\documentclass[seceq,preprint]{ptptex}

\preprintnumber[3cm]{
arXiv:0706.0409\\
KEK-TH-1156\\RIKEN-TH-102\\ June, 2007}

\title{
Comments on Solutions for Nonsingular Currents\\ 
in Open String Field Theories
}

\author{
Isao \textsc{Kishimoto}$^{1,}$\footnote{
E-mail:~ikishimo@riken.jp}
and
Yoji \textsc{Michishita}$^{2,}$\footnote{
E-mail:~michishi@post.kek.jp}
}

\inst{
$^{1}$Theoretical Physics Laboratory, RIKEN,\\ 
 Wako 351-0198, Japan\\
$^{2}$High Energy Accelerator Research Organization (KEK),\\
 Tsukuba 305-0801, Japan
}

\abst{
We investigate analytic solutions to Witten's bosonic string field theory and
 Berkovits' WZW-type superstring field theory.
We construct solutions with parameters out of simpler ones, using a
commutative monoid that includes the family of wedge states.
Our solutions are generalizations of solutions for marginal deformations by
nonsingular currents, and can also reproduce Schnabl's tachyon
vacuum solution in bosonic string field theory.
This implies that such known solutions are generated from 
simple solutions which are based on the identity state.
We also discuss gauge transformations and induced field redefinitions 
for our solutions in both bosonic and super string field theory.
}

\begin{document}

\maketitle


\section{Introduction}

The study of analytic solutions of string field theories is important
in order to understand nonperturbative phenomena in string theory.
There have been various attempts to solve the equations of motion 
of string field theories.
In particular, since Schnabl constructed an analytic
solution for tachyon condensation \cite{Schnabl_tach}
in Witten's cubic open string field theory,
there have been a number of new developments in this field.
Recently, new analytic solutions for marginal deformations
using nonsingular currents
have been proposed by Schnabl \cite{Schnabl:2007az}
and Kiermaier et al\cite{Kiermaier:2007ba}.
They can be regarded as an application of the technical methods
developed in Refs.~\citen{Schnabl_tach, RZpro} and \citen{ES}.
As a generalization of Refs.~\citen{Schnabl:2007az} and
\citen{Kiermaier:2007ba}
to Berkovits' WZW-type open superstring field theory,
new analytic solutions to the equation of motion
have also been constructed by Erler \cite{Erler_super}
and Okawa \cite{Okawa_super, Okawa_real}.

In this paper, we consider a generalization of 
$\Psi=\sqrt{|0\rangle}*(1+\hat{\psi}*A)^{-1}*
\hat{\psi}*\sqrt{|0\rangle}$,
which is the expression of the solution for marginal deformations given in
Ref.~\citen{Schnabl:2007az},
in both Witten's bosonic string field theory and
 Berkovits' WZW-type superstring field theory.
We construct solutions to the equation of motion
\begin{eqnarray}
\label{eq:EOM_bos}
 Q_{\rm B}\Psi+\Psi*\Psi=0
\end{eqnarray}
in bosonic string field theory and\footnote{
We often omit the symbol $*$ representing the star product for simplicity.}
\begin{eqnarray}
\label{eq:EOM_super}
\eta_0(e^{-\Phi}Q_{\rm B}e^{\Phi})=0
\end{eqnarray}
in superstring field theory
with parameters from rather simple ones, which we denote by $\hat{\psi}$
in bosonic string field theory and $\hat{\phi}$ in superstring field
theory, 
using a commutative monoid $\{P_{\alpha}\}_{\alpha\ge 0}$
with respect to the star product. A generalization of $A$ in Ref.~\citen{ES},
which connects any element $P_{\alpha}$ to the identity element, i.e.
$P_{\alpha=0}=I$ (the identity state), 
is also necessary in our method.
A simple and familiar example of a commutative monoid is 
the family of wedge states.
In bosonic string field theory, 
this $\hat{\psi}$ is BRST invariant and nilpotent, i.e.
$Q_{\rm B}\hat{\psi}=0$ and $\hat{\psi}*\hat{\psi}=0$.
Similarly, $\hat{\phi}$ satisfies $\eta_0Q_{\rm B}\hat\phi=0$,
$\hat{\phi}*\hat{\phi}=0$,
$\hat{\phi}*\eta_0\hat{\phi}=0$ and $\hat{\phi}*Q_{\rm B}\hat{\phi}=0$.
Explicit examples of $\hat{\psi}$ and $\hat{\phi}$
can be readily obtained using the identity state.
When we construct  $\hat{\psi}$ and $\hat{\phi}$
from (super) currents, the nilpotency (and
$\hat{\phi}*\eta_0\hat{\phi}=0$, $\hat{\phi}*Q_{\rm B}\hat{\phi}=0$)
 corresponds to 
the condition that the (super) currents be nonsingular.
If we use another $\hat{\psi}$, we obtain 
Schnabl's solution for tachyon condensation and its generalization.

The above facts imply that the recently developed solutions mentioned above
are all generated from the simple solutions $\hat{\psi}$ and $\hat{\phi}$,
 which are based on the identity state.
Also note that certain analytic solutions using the
identity state were studied before Schnabl presented the
solution in Ref.~\citen{Schnabl_tach}.
Considering these points, our observation regarding analytic solutions 
in bosonic and super string field theory
might be useful for the purpose of reconsidering 
previously derived solutions
and developing other new solutions.

The paper is organized as follows.
In \S \ref{sec:bosonic}, we construct
two-parameter solutions $\Psi^{(\alpha,\beta)}$
 to the equation of motion (\ref{eq:EOM_bos})
in bosonic string field theory constructed from $\hat{\psi}$,
 and then we
investigate marginal solutions and tachyon solutions by
choosing a particular  $\hat{\psi}$.
In \S \ref{sec:super}, noting the solutions in
Refs.~\citen{Erler_super} and \citen{Okawa_super},
we construct four types of solutions, $\Phi^{(\alpha,\beta)}_{(i)}$
($i=1,2,3,4$), to the equation of motion (\ref{eq:EOM_super})
in a manner analogous to that used in
 our method applied to bosonic string field theory.
We also comment on the reality condition
for the string field $\Phi$ \cite{Erler_super}
and solutions in path-ordered
form developed in Ref.~\citen{Okawa_real}.
In \S \ref{sec:gauge_tr}, we study gauge transformations
and induced field redefinitions for our solutions in both bosonic and
super string field theory. In \S \ref{sec:discussion}, we summarize
our results and discuss future directions.

\section{Solutions in bosonic string field theory
\label{sec:bosonic}}

Let us consider a family of BRST invariant 
string fields
${P_{\alpha}}$ ($\alpha\ge 0)$ with ghost number zero,
which satisfy
\begin{eqnarray}
&&Q_{\rm B}P_{\alpha}=0\,,~~~
P_{\alpha}*P_{\beta}=P_{\alpha+\beta}\,,~~~P_{\alpha=0}=I.
\label{eq:comm_monoid}
\end{eqnarray}
Here $I$ is the identity string field. 
Thus, such a family of string fields $\{P_{\alpha}\}_{\alpha\ge 0}$ 
forms a commutative monoid with respect to the star product.
In the following, we formally
 regard $I$ as the identity element with respect to
the star product of string fields.
We suppose that there exists a string field $A^{(\gamma)}$ 
for the above family satisfying
\begin{eqnarray}
 Q_{\rm B}A^{(\gamma)}&=&I-P_{\gamma}\,.
\label{eq:Agamma}
\end{eqnarray}
If we have a BRST invariant and nilpotent string field $\hat\psi$, i.e.
\begin{eqnarray}
\label{eq:psi_trivial}
&&Q_{\rm B}\hat{\psi}=0,~~~\hat{\psi}*\hat{\psi}=0\,,
\end{eqnarray}
with ghost number 1, then
\begin{eqnarray}
 \Psi^{(\alpha,\beta)}
&=&P_{\alpha}*\frac{1}{1+\hat{\psi}*A^{(\alpha+\beta)}}
*\hat{\psi}*P_{\beta}
\label{eq:b_sol_gen}
\end{eqnarray}
represents a solution to the equation of motion (\ref{eq:EOM_bos}).
Using the above assumption and the derivation property of the BRST operator
$Q_{\rm B}$, this can be proven as follows:
\begin{eqnarray}
 Q_{\rm B}\Psi^{(\alpha,\beta)}&=&P_{\alpha}*Q_{\rm B}\left(
\frac{1}{1+\hat{\psi}*A^{(\alpha+\beta)}}\right)*\hat\psi*P_{\beta}\nonumber\\
&=&-P_{\alpha}*
\frac{1}{1+\hat{\psi}*A^{(\alpha+\beta)}}
*(Q_{\rm B}(I+\hat{\psi}*A^{(\alpha+\beta)}))*
\frac{1}{1+\hat{\psi}*A^{(\alpha+\beta)}}
*\hat{\psi}*P_{\beta}\nonumber\\
&=&P_{\alpha}*
\frac{1}{1+\hat{\psi}*A^{(\alpha+\beta)}}
*\hat{\psi}*(Q_{\rm B}A^{(\alpha+\beta)})*
\frac{1}{1+\hat{\psi}*A^{(\alpha+\beta)}}
*\hat{\psi}*P_{\beta}\nonumber\\
&=&P_{\alpha}*
\frac{1}{1+\hat{\psi}*A^{(\alpha+\beta)}}
*\hat{\psi}*(I-P_{\alpha+\beta})*
\frac{1}{1+\hat{\psi}*A^{(\alpha+\beta)}}
*\hat{\psi}*P_{\beta}\nonumber\\
&=&P_{\alpha}*
\frac{1}{1+\hat{\psi}*A^{(\alpha+\beta)}}
*\hat{\psi}*\hat{\psi}*\frac{1}{1+A^{(\alpha+\beta)}*\hat{\psi}}
*P_{\beta}\nonumber\\
&&-P_{\alpha}*
\frac{1}{1+\hat{\psi}*A^{(\alpha+\beta)}}
*\hat{\psi}*P_{\beta}*P_{\alpha}*
\frac{1}{1+\hat{\psi}*A^{(\alpha+\beta)}}
*\hat{\psi}*P_{\beta}\nonumber\\
&=&-\Psi^{(\alpha,\beta)}*\Psi^{(\alpha,\beta)}\,.
\end{eqnarray}
The assumption concerning $\hat{\psi}$ given in (\ref{eq:psi_trivial}) 
implies that $\hat{\psi}$ itself satisfies the equation of motion
$Q_{\rm B}\hat{\psi}+\hat{\psi}*\hat{\psi}=0$
trivially. Therefore, we can also interpret (\ref{eq:b_sol_gen}) 
as representing a transformation from a rather trivial 
solution $\hat{\psi}$ to another solution.
If we replace $\hat{\psi}$ with $\lambda\hat\psi$ 
(where $\lambda$ is a parameter),
the condition (\ref{eq:psi_trivial}) is also satisfied.
In this sense, the new solution (\ref{eq:b_sol_gen}) obtained from a
BRST and nilpotent string field can naturally have a one-parameter
dependence.\footnote{\label{footnote:zeze}
Here, we should note that we can construct a solution to the equation of
motion using the formula (\ref{eq:b_sol_gen}) from a solution 
$\hat\psi$ that satisfies
$Q_{\rm B}\hat\psi+\hat\psi*\hat\psi=0$
instead of (\ref{eq:psi_trivial}).
This fact was pointed out by S.~Zeze.
The proof is almost the same as that above.
However, in that case, 
we cannot replace $\hat{\psi}$ with $\lambda\hat\psi$, in
general.
}

The most familiar example of commutative monoids (\ref{eq:comm_monoid}) 
is given by the wedge states,
\begin{eqnarray}
P_{\alpha}&=&U^{\dagger}_{\alpha+1}U_{\alpha+1}|0\rangle
=e^{-\frac{\alpha-1}{2}({\cal L}_0+{\cal L}_0^{\dagger})}|0\rangle
=e^{-\frac{\pi}{2}\alpha K_1^L}I\,,
\label{eq:wedgestates}
\end{eqnarray}
and correspondingly, we can take
\begin{eqnarray}
A^{(\gamma)}&=&\int_0^{\gamma}\!d\alpha\frac{\pi}{2}B_1^LP_{\alpha}\,,
\label{eq:Agamma_ex}
\end{eqnarray}
where we have used the notation of Ref.~\citen{Schnabl_tach}:
\begin{eqnarray}
&&{\cal B}_0=b_0+\sum_{k=1}^{\infty}\frac{2(-1)^{k+1}}{4k^2-1}b_{2k},
~~~{\cal B}_0^{\dagger}=b_0+\sum_{k=1}^{\infty}\frac{2(-1)^{k+1}}{
4k^2-1}b_{-2k},
\label{eq:B0_B0d}\\
&&B_1^L=\frac{1}{2}(b_1+b_{-1})+\frac{1}{\pi}({\cal B}_0
+{\cal B}_0^{\dagger}),\\
&&{\cal L}_0=\{Q_{\rm B},{\cal B}_0\},~~~~~
{\cal L}_0^{\dagger}=\{Q_{\rm B},{\cal B}_0^{\dagger}\},
~~~~~K_1^L=\{Q_{\rm B},B_1^L\}.
\end{eqnarray}
In Ref.~\citen{ES}, $A^{(\gamma=1)}=A$  plays 
an important role in the study of the vanishing cohomology around 
Schnabl's vacuum solution. 
If $\hat{\psi}$ satisfies the reality condition
${\rm bpz}^{-1}\!\circ{\rm hc}(\hat{\psi})=\hat{\psi}$,
where bpz and hc denote BPZ and Hermitian conjugation, respectively,
our solution  (\ref{eq:b_sol_gen}) with $\alpha=\beta\in {\mathbb R}$ 
also does, because 
${\rm bpz}^{-1}\!\circ{\rm hc}(A^{(\gamma)})=A^{(\gamma)}$
for $\gamma\in {\mathbb R}$.
We could also consider other families satisfying (\ref{eq:comm_monoid})
and (\ref{eq:Agamma}) using appropriate conformal frames studied in
Refs.~\citen{RZpro} and \citen{Okawa:2006sn}.

As an example of (\ref{eq:psi_trivial}), let us consider
\begin{eqnarray}
 \hat\psi&=&\lambda_{\rm s}\hat{\psi}_{\rm s}+\lambda_{\rm
  m}\hat{\psi}_{\rm m}\,,
\label{eq:psi_sm}\\
\hat{\psi}_{\rm s}&=&Q_{\rm B}\hat{\Lambda}_0\,,~~~~
\hat{\Lambda}_0\equiv U_1^{\dagger}U_1B_1^Lc_1|0\rangle\,,\\
\hat{\psi}_{\rm m}&=&U_1^{\dagger}U_1c J(0)|0\rangle\,,
\end{eqnarray}
where $J(z)$ is a matter primary field of dimension 1 
such that $J(y)J(z)$ is not singular for $y\to z$.
In the case $J=\zeta_aJ^a$, with coefficients $\zeta_a$,
where $J^a$ satisfies the OPE
\begin{eqnarray}
\label{eq:JJ_OPE}
 J^a(y)J^b(z)&\sim&\frac{g^{ab}}{(y-z)^2}+\frac{1}{y-z}if^{ab}_{~~c}J^c(z)
+\cdots
\end{eqnarray}
(with $f^{ab}_{~~c}$ being the structure constant of the associated Lie
algebra), the nonsingular condition for $J$ 
is $\zeta_a\zeta_b g^{ab}=0$. The quantities
$\lambda_{\rm s}$ and $\lambda_{\rm m}$ in (\ref{eq:psi_sm}) are
parameters.

In fact, the  BRST invariance of $\hat\psi_{\rm s}$ and $\hat\psi_{\rm
m}$ follows trivially.
The nilpotency of $\hat\psi_{\rm m}$ follows from\footnote{
In the following computations, we often use the 
following star product formula in the sliver frame:
\begin{eqnarray}
&&U_r^{\dagger}U_r\tilde{\phi}_1(\tilde{x}_1)\cdots 
\tilde{\phi}_n(\tilde{x}_n)|0\rangle *
U_s^{\dagger}U_s\tilde{\psi}_1(\tilde{y}_1)\cdots 
\tilde{\psi}_m(\tilde{y}_m)|0\rangle\\
&=&U_{r+s-1}^{\dagger}U_{r+s-1}
\tilde{\phi}_1(\tilde{x}_1\!+\!\frac{\pi}{4}(s-1))\cdots 
\tilde{\phi}_n(\tilde{x}_n\!+\!\frac{\pi}{4}(s-1))\tilde{\psi}_1
(\tilde{y}_1\!-\!\frac{\pi}{4}(r-1))\cdots 
\tilde{\psi}_m(\tilde{y}_m\!-\!\frac{\pi}{4}(r-1))|0\rangle.\nonumber
\end{eqnarray}
This formula is given in Ref.~\citen{Schnabl_tach}.
}
\begin{eqnarray}
 \hat\psi_{\rm m}*\hat\psi_{\rm m}=U_1^{\dagger}U_1\tilde c
\tilde J(0)|0\rangle*U_1^{\dagger}U_1\tilde c
\tilde J(0)|0\rangle
=U_1^{\dagger}U_1
\tilde c\tilde J(0)\tilde c\tilde J(0)|0\rangle\,,
\end{eqnarray}
where $\tilde c(\tilde z)$ and $ \tilde J(\tilde z)$ are fields on 
the semi-infinite cylinder $C_{\pi}$ 
(the sliver frame) obtained with the conformal map $\tilde z=\arctan z$
from the conventional upper half plane, and we have $cJ(0)cJ(0)=0$,
which results from the nonsingular condition of the current $J$.
Noting the relation
\begin{eqnarray}
&&Q_{\rm B}\hat{\Lambda}_0*\hat{\Lambda}_0=K_1^L\hat{\Lambda}_0
*\hat{\Lambda}_0
-U_1^{\dagger}U_1B_1^L\tilde c\tilde\partial\tilde c(0)|0\rangle
*\hat{\Lambda}_0\nonumber\\
&&=(K_1^LB_1^RU_1^{\dagger}U_1c_1|0\rangle)
*U_1^{\dagger}U_1c_1|0\rangle
-(B_1^RB_1^LU_1^{\dagger}U_1
\tilde c\tilde\partial\tilde c(0)|0\rangle)*
U_1^{\dagger}U_1c_1|0\rangle\nonumber\\
&&=K_1^L(B_1U_1^{\dagger}U_1c_1|0\rangle
*U_1^{\dagger}U_1c_1|0\rangle)-
K_1^LB_1^L(U_1^{\dagger}U_1\tilde c(0)|0\rangle
*U_1^{\dagger}U_1\tilde c(0)|0\rangle)\nonumber\\
&&~~~+B_1^L(U_1^{\dagger}U_1\tilde
\partial\tilde c(0)|0\rangle)*
U_1^{\dagger}U_1\tilde c(0)|0\rangle)
\nonumber\\
&&=K_1^LU_1^{\dagger}U_1c_1|0\rangle-B_1^LU_1^{\dagger}U_1
\tilde c\tilde\partial\tilde c(0)|0\rangle
=Q_{\rm B}\hat{\Lambda}_0\,,
\label{eq:QLLQL}
\end{eqnarray}
with the derivation property of $Q_{\rm B}$
and nilpotency, we have $Q_{\rm B}\hat{\Lambda}_0*Q_{\rm
B}\hat{\Lambda}_0=0$.
Similarly, we find
\begin{eqnarray}
\hat{\psi}_{\rm m}*\hat{\psi}_{\rm s}&=&-Q_{\rm B}(\hat{\psi}_{\rm
  m}*\hat{\Lambda}_0)=-Q_{\rm B}(\hat{\psi}_{\rm
  m}*(|I\rangle-B_1^RU_1^{\dagger}U_1c_1|0\rangle))\nonumber\\
&=&-Q_{\rm B}B_1^R(\hat{\psi}_{\rm m}*U_1^{\dagger}U_1\tilde
c(0)|0\rangle)
=-Q_{\rm B}B_1^RU_1^{\dagger}U_1\tilde c\tilde J(0)
\tilde c(0)|0\rangle=0\,,\\
\hat{\psi}_{\rm s}*\hat{\psi}_{\rm m}&=&Q_{\rm
 B}(\hat{\Lambda}_0*\hat{\psi}_{\rm
  m})=Q_{\rm B}B_1^L(U_1^{\dagger}U_1\tilde
  c(0)|0\rangle*\hat{\psi}_{\rm m})\nonumber\\
&=&Q_{\rm B}B_1^LU_1^{\dagger}U_1\tilde c(0)
\tilde c\tilde J(0)|0\rangle=0\,.
\end{eqnarray}
Therefore, we conclude that $\hat{\psi}$ is 
BRST invariant and nilpotent
for any values of $\lambda_{\rm s}$ and $\lambda_{\rm m}$.
We note that 
$\hat\psi_{\rm m}$ and $\hat\psi_{\rm s}$ are ``identity-based''
solutions in the sense that $U_1^{\dagger}U_1=e^{{\frac{1}{2}}({\cal
L}_0+{\cal L}_0^{\dagger})}$ yields the identity state:
$I=U_1^{\dagger}U_1|0\rangle$.

Although there exist other solutions to (\ref{eq:psi_trivial}),
such as $\hat{\psi}=Q_L I$ \cite{Horowitz:1986dt},
we explicitly examine the solutions $\Psi^{(\alpha,\beta)}$ 
appearing in (\ref{eq:b_sol_gen})
generated from $\hat{\psi}$ given by (\ref{eq:psi_sm})
in the cases $\lambda_{\rm s}=0$ and $\lambda_{\rm m}=0$.
In the following, we take the family of wedge states as
$\{P_{\alpha}\}_{\alpha\ge 0}$ for simplicity.

\subsection{Solution for marginal deformation}

First, we consider the case $\lambda_{\rm s}=0$ in (\ref{eq:psi_sm}).
The solution (\ref{eq:b_sol_gen}) is expanded as
\begin{eqnarray}
 \Psi^{(\alpha,\beta)}&=&
\sum_{k=0}^{\infty}(-1)^k\lambda_{\rm m}^{k+1}
P_{\alpha}*(\hat{\psi}_{\rm m}*A^{(\alpha+\beta)})^k*\hat{\psi}_{\rm
m}*P_{\beta}
=\sum_{n=1}^{\infty}\lambda_{\rm m}^n\psi_{{\rm m},n}\,,
\label{eq:rs_boson}
\end{eqnarray}
where
\begin{eqnarray}
\psi_{{\rm m},1}&=&U^{\dagger}_{\alpha+\beta+1}U_{\alpha+\beta+1}
\tilde c\tilde J\biggl(\frac{\pi}{4}(\beta-\alpha)\biggr)|0\rangle\,,\\
\psi_{{\rm
 m},k+1}
&=&\left(-\frac{\pi}{2}\right)^k\int_0^{\alpha+\beta}dr_1
\cdots\int_0^{\alpha+\beta}dr_k\,
U^{\dagger}_{\alpha+\beta+1
+\sum_{l=1}^kr_l}U_{\alpha+\beta+1+\sum_{l=1}^kr_l}\nonumber\\
&&\times 
\prod_{m=0}^k\tilde J\biggl(\frac{\pi}{4}
(\beta-\alpha-\sum_{l=1}^mr_l+\sum_{l=m+1}^kr_l)\biggr)\\
&&\times \biggl[-\frac{1}{\pi}\hat{\cal B}\tilde
 c\biggl(\frac{\pi}{4}\biggl(\beta-\alpha+\sum_{l=1}^kr_l\biggr)\biggr)
\tilde c\biggl(\frac{\pi}{4}\biggl(\beta-\alpha-\sum_{l=1}^kr_l
\biggr)\biggr)\nonumber\\
&&~~~~+\frac{1}{2}\biggl\{
\tilde c\biggl(\frac{\pi}{4}\biggl(\beta-\alpha+\sum_{l=1}^kr_l
\biggr)\biggr)
+\tilde
 c\biggl(\frac{\pi}{4}\biggl(\beta-\alpha-\sum_{l=1}^kr_l
\biggr)\biggr)\biggr\}
\biggr]|0\rangle\,.\nonumber
\end{eqnarray}
The quantities $\psi_{m,k}$ are characterized as follows:
\begin{eqnarray}
\label{eq:solo1}
&&Q_{\rm B}\psi_{{\rm m},1}=0\,,~~~~~{\cal B}^{(\alpha,\beta)}
\psi_{{\rm m},1}=0\,,\\
&&\psi_{{\rm m},k+1}
=-\frac{{\cal B}^{(\alpha,\beta)}}{{\cal L}^{(\alpha,\beta)}}
\sum_{l=1}^{k}\psi_{{\rm m},l}*\psi_{{\rm m},k-l+1}\,,
\label{eq:solok+1}
\end{eqnarray}
where
\begin{eqnarray}
\label{eq:gen_Brs}
 {\cal B}^{(\alpha,\beta)}&=&\frac{1}{2}(\alpha+\beta-1)\hat{\cal B}
+{\cal B}_0+\frac{\pi}{4}(\alpha-\beta)B_1\,,\\
&&\hat{\cal B}\equiv{\cal B}_0+{\cal B}_0^{\dagger}\,,~~~
B_1=b_1+b_{-1}\,,\\
 {\cal L}^{(\alpha,\beta)}&\equiv&\{Q_{\rm B},{\cal B}^{(\alpha,\beta)}\}=
\frac{1}{2}(\alpha+\beta-1)\hat{\cal L}+{\cal L}_0
+\frac{\pi}{4}(\alpha-\beta)K_1\,,\\
&&\hat{\cal L}\equiv{\cal L}_0+{\cal L}_0^{\dagger}\,,~~~
K_1=L_1+L_{-1}\,.
\end{eqnarray}
Here we have used the identities
\begin{eqnarray}
&&e^{t {\cal
  L}^{(\alpha,\beta)}}=e^{\frac{1}{2}(e^t-1)(\alpha+\beta-1)\hat{\cal
  L}}\,
e^{t{\cal L}_0}\,e^{\frac{\pi}{4}(\alpha-\beta)(1-e^{-t})K_1},\\
&&\frac{{\cal B}^{(\alpha,\beta)}}{{\cal L}^{(\alpha,\beta)}}
={\cal B}^{(\alpha,\beta)}\int_0^{\infty}dT\,e^{-T {\cal
L}^{(\alpha,\beta)}},\\
&&e^{-T {\cal L}^{(\alpha,\beta)}}U^{\dagger}_{\gamma}U_{\gamma}=
U^{\dagger}_{(\gamma-\alpha-\beta-1)e^{-T}+\alpha+\beta+1}
U_{(\gamma-\alpha-\beta-1)e^{-T}+\alpha+\beta+1}
e^{-T{\cal L}_0}e^{\frac{\pi}{4}(\alpha-\beta)(1-e^T)K_1},
\nonumber\\
\\
&&{\cal B}^{(\alpha,\beta)}U^{\dagger}_{\gamma}U_{\gamma}=
U^{\dagger}_{\gamma}U_{\gamma}\left(
\frac{1}{2}(\alpha+\beta+1-\gamma)\hat{\cal B}+{\cal B}_0
+\frac{\pi}{4}(\alpha-\beta)B_1\right)\,,\\
&&\{{\cal B}_0,\tilde c(\tilde z)\}=\tilde z\,,~~~~~
\{B_1,\tilde c(\tilde z)\}=1\,,\\
&&U_{\beta}\tilde \phi(\tilde z)U_{\beta}^{-1}
=(2/\beta)^{h_{\phi}}\tilde \phi(
(2/\beta)\tilde z)\,,~~~~~
e^{a K_1}\tilde \phi(\tilde z)e^{-aK_1}=\tilde \phi(\tilde z+a)
\end{eqnarray}
to check the above relations. 
We thus find that the solution $\Psi^{(\alpha,\beta)}$
given in (\ref{eq:rs_boson})
satisfies the ``generalized Schnabl gauge'' condition,\footnote{
In the case $\alpha=\beta\to \infty$ ,
this gauge condition becomes 
``the modified Schnabl gauge'' used in Ref.~\citen{FNS}.
}
\begin{eqnarray}
\label{eq:b_gauge_cond}
 {\cal B}^{(\alpha,\beta)}\Psi^{(\alpha,\beta)}=0\,,
\end{eqnarray}
and at each order with respect to the parameter 
$\lambda_{\rm m}$, we have
\begin{eqnarray}
&&Q_{\rm B}\psi_{{\rm m},1}=0\,,~~~~
Q_{\rm B}\psi_{{\rm m},k+1}=-\sum_{l=1}^{k}\psi_{{\rm
m},l}*\psi_{{\rm m},k-l+1}\,,
\label{eq:EOM_order}
\end{eqnarray}
{}from (\ref{eq:solo1}) and (\ref{eq:solok+1}),
as noted in Ref.~\citen{Michishita:2005se}.
Then, using the formula
\begin{eqnarray}
P_{\alpha-\frac{1}{2}}*\psi*P_{\beta-\frac{1}{2}}
=e^{(\beta-\frac{1}{2})\frac{\pi}{2}K_1^R
-(\alpha-\frac{1}{2})\frac{\pi}{2}K_1^L}\psi
=e^{\frac{\pi}{4}(\beta-\alpha)K_1}e^{-\frac{1}{2}(\alpha+\beta-1)\hat{\cal
L}}\psi,
\end{eqnarray}
{}from Ref.~\citen{Schnabl_tach}, and the relations
\begin{eqnarray}
&&e^{t\hat{\cal L}}=(1-2t)^{\frac{D}{2}}(1-2t)^{-{\cal L}_0}\,,
~~~~
D\equiv{\cal L}_0-{\cal L}_0^{\dagger}\,,\\
&&\psi_{{\rm m},k+1}=\nonumber\\
&&\left(-\frac{\pi}{2}\right)^k\int_0^{\alpha+\beta}\!dr_1
\cdots\int_0^{\alpha+\beta}\!dr_kP_{\alpha-\frac{1}{2}}*
\prod_{m=1}^k(cJ(0)|0\rangle*B_1^L|r_m\rangle)*cJ(0)|0\rangle
*P_{\beta-\frac{1}{2}},\nonumber\\
\end{eqnarray}
the solution $\Psi^{(\alpha,\beta)}$ can be rewritten as
\begin{eqnarray}
 \Psi^{(\alpha,\beta)}=e^{\frac{\pi}{4}(\beta-\alpha)K_1}
(\alpha+\beta)^{\frac{D}{2}}
\Psi^{(1/2,1/2)},
\label{eq:rs_derivation}
\end{eqnarray}
for $\alpha+\beta>0$,
where both $K_1=L_1+L_{-1}$ and $D= {\cal L}_0-{\cal L}_0^{\dagger}$ are
derivations with respect to the star product and are BPZ odd.
The solution $\Psi^{(1/2,1/2)}$ appearing here is investigated in
Refs.~\citen{Schnabl:2007az} and \citen{Kiermaier:2007ba}.
The relation (\ref{eq:rs_derivation}) itself 
seems to be singular in the limit $\alpha+\beta\to 0$, although
$\Psi^{(0,0)}=\lambda_{\rm m}U_1^{\dagger}U_1cJ(0)|0\rangle
=\hat{\psi}|_{\lambda_{\rm s}=0}$ 
is a BRST invariant and nilpotent solution.
We note the identity
\begin{eqnarray}
 {\cal B}^{(\alpha,\beta)}e^{\frac{\pi}{4}(\beta-\alpha)K_1}
(\alpha+\beta)^{\frac{D}{2}}
=e^{\frac{\pi}{4}(\beta-\alpha)K_1}(\alpha+\beta)^{\frac{D}{2}}{\cal
B}_0\,,
\end{eqnarray}
which is consistent with (\ref{eq:b_gauge_cond}).

Let us consider the overlapping of $\tilde\varphi(0)c_1c_0|0\rangle$
and the solution $\Psi^{(\alpha,\beta)}$,
where $\varphi$ is a matter primary field of dimension $h_{\varphi}$.
Then, using
\begin{eqnarray}
&&\langle 0|c_{-1}c_0\left[-\frac{1}{\pi}\hat{\cal B}\tilde c(\tilde x_+)
\tilde c(\tilde x_-)+\frac{1}{2}(\tilde c(\tilde x_+)+
\tilde c(\tilde x_-))
\right]|0\rangle\nonumber\\
&&=\frac{1}{2}\cos^2\tilde x_+\left[
1+\frac{1}{\pi}(2\tilde x_-+\sin(2\tilde x_-))
\right]
+\frac{1}{2}\cos^2\tilde x_-\left[
1-\frac{1}{\pi}(2\tilde x_++\sin(2\tilde x_+))
\right]\nonumber\\
\end{eqnarray}
in the ghost sector with 
the normalization $\langle 0|c_{-1}c_0c_1|0\rangle=1$,
we have
\begin{eqnarray}
 &&\langle
  0|c_{-1}c_0\,I\circ\tilde\varphi(0)|\psi_{{\rm m},k+1}\rangle\nonumber\\
&&=\left(-\frac{\pi}{2}\right)^k
\int_0^{\alpha+\beta}dr_1\cdots\int_0^{\alpha+\beta}dr_k
\left(\frac{2}{\alpha+\beta+1+\sum_{l=1}^k\!r_l}\right)^{k-1+h_{\varphi}}
\nonumber\\
&&\times \biggl[
\frac{1}{2}\cos^2\!\frac{\pi(\beta-\alpha+\sum_{l=1}^kr_l)}{
2(\alpha+\beta+1+\sum_{l=1}^k\!r_l)}
\biggl(
\frac{2\beta+1}{\alpha+\beta+1+\sum_{l=1}^k\!r_l}\!-\!\frac{1}{\pi}
\sin\frac{\pi(2\beta+1)}{\alpha+\beta+1+\sum_{l=1}^k\!r_l}
\biggr)\nonumber\\
&&
+\frac{1}{2}\cos^2\!\frac{\pi(\beta-\alpha-\sum_{l=1}^k\!r_l)}{
2(\alpha+\beta+1+\sum_{l=1}^k\!r_l)}
\biggl(
\frac{2\alpha+1}{\alpha+\beta+1+\sum_{l=1}^k\!r_l}\!-\!\frac{1}{\pi}
\sin\frac{\pi(2\alpha+1)}{\alpha+\beta+1+\sum_{l=1}^k\!r_l}
\biggr)
\biggr]\nonumber\\
&&\times\left\langle I\circ\tilde \varphi(0)
\prod_{m=0}^k
\lambda_{a_m}\tilde J^{a_m}\biggl(\frac{\pi
(\beta-\alpha-\sum_{l=1}^mr_l+\sum_{l=m+1}^kr_l)}{
2(\alpha+\beta+1+\sum_{l=1}^k\!r_l)}
\biggr)\right\rangle\,.
\label{eq:c1coeff}
\end{eqnarray}
Here $I$ is the inversion map, which becomes $I(\tilde z)=\tilde z\pm
\pi/2$ on the sliver frame $C_{\pi}$, and
the last factor is the matter correlator,
which is evaluated as
\begin{eqnarray}
\label{eq:mat_corr}
&&\left\langle I\circ\tilde \varphi(0)
\prod_{m=0}^k
\tilde J(\tilde x_m)\right\rangle=
\left\langle I\circ\varphi(0)
\prod_{m=0}^k(\cos\tilde x_m)^{-2}
J(\tan\tilde x_m)\right\rangle_{{\rm UHP}}\,,\\
&&\tilde x_m \equiv \frac{\pi
(\beta-\alpha-\sum_{l=1}^mr_l+\sum_{l=m+1}^kr_l)}{
2(\alpha+\beta+1+\sum_{l=1}^kr_l)}\,.
\end{eqnarray}
In the following, we perform explicit calculations for some coefficients
of the solutions, which have $(\alpha,\beta)$ or gauge dependence.
\\

\noindent[Rolling tachyon]~
Firstly, we consider the case $J= :\!e^{X^0}\!\!:$,
which is dimension one primary and non-singular because
of the OPE:
\begin{eqnarray}
 :\!e^{nX^0(x)}\!\!::\!e^{mX^0(y)}\!\!:\simeq |x-y|^{2nm}
:\!e^{(n+m)X^0(y)}\!\!:\,.
\end{eqnarray}
Then, the matter correlator (\ref{eq:mat_corr}) for 
$\varphi=:\!e^{-(k+1)X^0}\!\!:$ is evaluated as
\begin{eqnarray}
&&\langle I\circ\!:\!e^{-(k+1)\tilde X^0(0)}\!\!:
:\!e^{\tilde X^0(\tilde x_0)}\!\!:
:\!e^{\tilde X^0(\tilde x_1)}\!\!:
\cdots
\!:\!e^{\tilde X^0(\tilde x_k)}\!\!:\rangle/V_{26}\nonumber\\
&&=\prod_{m=0}^k(\cos\tilde x_m)^{-2}\!\!
\prod_{0\le i<j\le k }\!\!|\tan\tilde x_i-\tan\tilde x_j|^2
=\prod_{m=0}^k(\cos\tilde x_m)^{-2(k+1)}\!\!
\prod_{0\le i<j\le k }\!\!\sin^2(\tilde x_i-\tilde x_j).\nonumber\\
\end{eqnarray}
Using this formula and $h_{\phi}=(k+1)^2$ for 
$\varphi=:\!e^{-(k+1)X^0}\!\!:$, we compute (\ref{eq:c1coeff})
for certain values of $k$ as follows.
In the case $k=0$, we have
\begin{eqnarray}
\langle 0|c_{-1}c_0 I\circ\!:\!e^{-\tilde X^0(0)}\!\!:|
\psi_{{\rm m},1}\rangle/V_{26}=1
\end{eqnarray}
for all $\alpha,\beta$. For $k=1$ and $\alpha=\beta$, we have
\begin{eqnarray}
&&\langle 0|c_{-1}c_0 I\circ\!:\!e^{-2\tilde X^0(0)}\!\!:
|\psi_{{\rm m},2}\rangle/V_{26}\nonumber\\
&&=\frac{-\pi}{2}\int_0^{2\alpha} dx\left(
\frac{2}{2\alpha+1+x}
\right)^4\frac{\sin^2\frac{\pi(2\alpha+1)}{2\alpha+1+x}}{\sin^6
\frac{\pi(2\alpha+1)}{2(2\alpha+1+x)}}\left(
\frac{2\alpha+1}{2\alpha+1+x}-\frac{1}{\pi}\sin
\frac{\pi(2\alpha+1)}{2\alpha+1+x}
\right)\nonumber\\
&&=-\frac{64\cot^3\frac{\pi(2\alpha+1)}{2(4\alpha+1)}}{
3(4\alpha+1)^3},
\end{eqnarray}
which  is equal to $
-\frac{64}{243\sqrt{3}}$ 
for $\alpha=1/2$, \cite{Schnabl:2007az, Kiermaier:2007ba}
as expected. For $k=2,\alpha=\beta$, we have
\begin{eqnarray}
&&\langle 0|c_{-1}c_0 I\circ\!:\!e^{-3\tilde X^0(0)}\!\!:|
\psi_{{\rm m},3}\rangle/V_{26}=\nonumber\\
&&\frac{\pi^2}{4}\!\int_0^{2\alpha}\!dx\!
\int_0^{2\alpha}\!dy\!\left(\!
\frac{2}{(2\alpha+1+x+y)
\sin\frac{\pi(2\alpha+1)}{2(2\alpha+1+x+y)}}
\!\right)^{10}
\left(\!\sin\!\frac{\pi(2\alpha+1+2x)}{2(2\alpha+1+x+y)}
\right)^{-6}\nonumber\\
&&~~~~~~~~~~~~~~~
\times \left(
\frac{2\alpha+1}{2\alpha+1+x+y}-\frac{1}{\pi}\sin
\frac{\pi(2\alpha+1)}{2\alpha+1+x+y}
\right)\\
&&~~~~~~~~~~~~~~~
\times 
\left(
\sin\frac{\pi(2\alpha+1)}{2\alpha+1+x+y}
\sin\frac{\pi x}{2\alpha+1+x+y}
\sin\frac{\pi y}{2\alpha+1+x+y}
\right)^2,\nonumber
\end{eqnarray}
which is numerically evaluated as $0.00214766$
in the case
 $\alpha=1/2$\cite{Schnabl:2007az, Kiermaier:2007ba}.
For $k\ge 1$, (\ref{eq:c1coeff}) can be rewritten as
\begin{eqnarray}
 &&\langle
  0|c_{-1}c_0\,I\circ \!:\!e^{-(k+1)X^0(0)}\!\!:|
\psi_{{\rm m},k+1}\rangle/V_{26}\nonumber\\
&&=\left(-\frac{\pi}{2}\right)^k(\alpha+\beta)^{-k^2-2k}
\int_0^1dx_1\cdots\int_0^1dx_k
\left(\frac{2}{1+\frac{1}{\alpha+\beta}+\sum_{l=1}^kx_l}\right)^{k^2+3k}
\nonumber\\
&&\times \biggl[
\frac{1}{2}\sin^2\frac{\pi(\frac{2\beta+1}{\alpha+\beta}+2\sum_{l=1}^kx_l)}{
2(1+\frac{1}{\alpha+\beta}+\sum_{l=1}^kx_l)}
\biggl(
\frac{\frac{2\beta+1}{\alpha+\beta}}{1+\frac{1}{\alpha+\beta}
+\sum_{l=1}^kx_l}\!-\!\frac{1}{\pi}
\sin\frac{\pi\frac{2\beta+1}{\alpha+\beta}}{1+\frac{1}{\alpha+\beta}
+\sum_{l=1}^kx_l}\biggr)\nonumber\\
&&
+\frac{1}{2}\sin^2\frac{\pi(\frac{2\alpha+1}{\alpha+\beta}
+2\sum_{l=1}^kx_l)}{2(1+\frac{1}{\alpha+\beta}
+\sum_{l=1}^kx_l)}\biggl(
\frac{\frac{2\alpha+1}{\alpha+\beta}}{1+
\frac{1}{\alpha+\beta}+\sum_{l=1}^kx_l}\!-\!\frac{1}{\pi}
\sin\frac{\pi\frac{2\alpha+1}{\alpha+\beta}}{1
+\frac{1}{\alpha+\beta}+\sum_{l=1}^kx_l}\biggr)\biggr]\nonumber\\
&&\times\left[\prod_{m=0}^k\sin^{-2(k+1)}
\frac{\pi(\frac{2\alpha+1}{\alpha+\beta}+2\sum_{l=1}^mx_l)}{
2(1+\frac{1}{\alpha+\beta}+\sum_{l=1}^kx_l)}\right]
\prod_{0\le i<j\le k }\!\sin^2\!\frac{\pi\sum_{l=i+1}^jx_l}{
2(1+\frac{1}{\alpha+\beta}+\sum_{l=1}^kx_l)},
\end{eqnarray}
which behaves as $(\alpha+\beta)^{-k^2-2k}$, up to constant factor,
in the case $\alpha+\beta\to \infty$.
\\

\noindent[Light-cone deformation]~
In the case $J=i\partial X^+$
and $\varphi=-\frac{1}{2}\partial X^-\partial X^-$,
with the OPE $\partial X^-(y)\partial X^+(z)\sim (y-z)^{-2}$,
the matter contribution (\ref{eq:mat_corr})
for $k=1$ is
\begin{eqnarray}
&&\langle I\circ\tilde\varphi(0) 
i\partial X^+(\tilde x_0)i\partial X^+(\tilde x_1)\rangle/V_{26}
=(\cos\tilde x_0\cos\tilde x_1)^{-2}.
\end{eqnarray}
Therefore, (\ref{eq:c1coeff}) is computed as
\begin{eqnarray}
&&\langle 0|c_{-1}c_0I\circ\tilde\varphi(0)|\psi_{{\rm m},2}
\rangle/V_{26}\nonumber\\
&&=\frac{-\pi}{2}\int_0^{\alpha+\beta}\!dx\!\left(\frac{2}{\alpha+\beta
+1+x}\right)^2\nonumber\\
&&~~~\times\biggl[
\frac{1}{2\sin^2\frac{\pi(2\beta+1)}{2(\alpha+\beta+1+x)}}
\left(\frac{2\beta+1}{\alpha+\beta+1+x}
-\frac{1}{\pi}\sin\frac{\pi(2\beta+1)}{\alpha+\beta+1+x}
\right)\nonumber\\
&&~~~+\frac{1}{2\sin^2\frac{\pi(2\alpha+1)}{2(\alpha+\beta+1+x)}}
\left(
\frac{2\alpha+1}{\alpha+\beta+1+x}
-\frac{1}{\pi}\sin\frac{\pi(2\alpha+1)}{\alpha+\beta+1+x}
\right)\biggr]\nonumber\\
&&=-\frac{4\cot\frac{\pi(2\alpha+1)}{2(4\alpha+1)}}{4\alpha+1}~~~~
({\rm for}~\alpha=\beta)\,,
\end{eqnarray}
and this gives $-4/(3\sqrt{3})$
for  $\alpha=\beta=1/2$, as in Ref.~\citen{Kiermaier:2007ba}.

\subsection{Tachyon solution}

Next, we consider the case $\lambda_{\rm m}=0$ in (\ref{eq:psi_sm}).
In this case, the solution is expanded as
\begin{eqnarray}
\Psi^{(\alpha,\beta)}&=&\sum_{k=0}^{\infty}(-1)^k
\lambda_{\rm s}^{k+1}P_{\alpha}*\hat{\psi}_{\rm s}*(A^{(\alpha+\beta)}
*\hat{\psi}_{\rm s})^k*P_{\beta}
=\sum_{n=1}^{\infty}\lambda_{\rm s}^n\psi_{{\rm s},n}\,.
\label{eq:psi_s_ab}
\end{eqnarray}
Then, noting (\ref{eq:QLLQL}) and
\begin{eqnarray}
\hat{\Lambda}_0*\hat{\Lambda}_0&=&-(B_1^R\hat{\Lambda}_0)
*U_1^{\dagger}U_1c_1|0\rangle=(B_1^LB_1U_1^{\dagger}U_1c_1|0\rangle)
*U_1^{\dagger}U_1c_1|0\rangle\nonumber\\
&=&B_1^L|I\rangle*U_1^{\dagger}U_1c_1|0\rangle
=\hat{\Lambda}_0\,,\\
A^{(\alpha+\beta)}*\hat{\psi}_{\rm s}
&=&\int_0^{\alpha+\beta}d\gamma\,\frac{\pi}{2}B_1^LP_{\gamma}*
\hat{\psi}_{\rm s}=
\int_0^{\alpha+\beta}d\gamma\,\frac{\pi}{2}(-B_1^RP_{\gamma})*
Q_{\rm B}\hat{\Lambda}_0\nonumber\\
&=&\int_0^{\alpha+\beta}d\gamma\,\frac{\pi}{2}P_{\gamma}*
B_1^LQ_{\rm B}\hat{\Lambda}_0
=\int_0^{\alpha+\beta}d\gamma\,\frac{\pi}{2}P_{\gamma}*
K_1^L\hat{\Lambda}_0\nonumber\\
&=&-\int_0^{\alpha+\beta}d\gamma\,\frac{\pi}{2}K_1^RP_{\gamma}
*\hat{\Lambda}_0
=\hat{\Lambda}_0-P_{\beta}*P_{\alpha}*\hat{\Lambda}_0\,,
\end{eqnarray}
the ${\cal O}(\lambda_{\rm s}^n)$ term of $\Psi^{(\alpha,\beta)}$ 
can be rewritten as
\begin{eqnarray}
\psi_{{\rm s},n}&=&P_{\alpha}*(Q_{\rm
B}\hat{\Lambda}_0)*P_{\beta}*(
P_{\alpha}*\hat{\Lambda}_0*P_{\beta}-I)^{n-1}
\nonumber\\
&=&-\sum_{l=0}^{n-1}\frac{(-1)^{n-1-l}(n-1)!}{l!(n-1-l)!}
\partial_t\psi_{t,l}^{(\alpha,\beta)}|_{t=0}\,,
\label{eq:rel_psis}
\\
\psi_{t,n}^{(\alpha,\beta)}
&=&\frac{2}{\pi}U_{n(\alpha+\beta)+t+\alpha+\beta+1}^{\dagger}
U_{n(\alpha+\beta)+t+\alpha+\beta+1}
\biggl[\nonumber\\
&&-\frac{1}{\pi}\hat{\cal B}
\tilde c\biggl(\frac{\pi}{4}(\beta-\alpha+t+n(\alpha+\beta))\!\biggr)
\tilde c\biggl(\frac{\pi}{4}(\beta-\alpha-t-n(\alpha+\beta))\!\biggr)
\\
&&+\frac{1}{2}\biggl\{
\tilde c\biggl(\frac{\pi}{4}(\beta-\alpha+t+n(\alpha+\beta))\!\biggr)
\!+
\tilde c\biggl(\frac{\pi}{4}(\beta-\alpha-t-n(\alpha+\beta))\!\biggr)
\biggr\}
\biggr]|0\rangle,\nonumber
\end{eqnarray}
where use has been made of the relations
\begin{eqnarray}
&&(P_{\alpha}*\hat\Lambda_0*P_{\beta})^n
=U_{n(\alpha+\beta)+1}^{\dagger}U_{n(\alpha+\beta)+1}B_1^L
\tilde c\biggl(\frac{\pi}{4}(2\beta-n(\alpha+\beta))\biggr)|0\rangle,\\
&&~~~~~~~~~~~~~~~~~~~~~~~~~~~~~~~
(n=1,2,\cdots),\nonumber\\
&&P_{\alpha}*Q_{\rm B}\hat\Lambda_0*P_{\beta}
*(P_{\alpha}*\hat\Lambda_0*P_{\beta})^n=
-\partial_t\psi_{t,n}^{(\alpha,\beta)}|_{t=0}\,,~~~~(n=0,1,2,\cdots).
\label{eq:QBLambda0}
\end{eqnarray}
Therefore, (\ref{eq:psi_s_ab}) can be re-summed as
\begin{eqnarray}
\Psi^{(\alpha,\beta)}&=&-\sum_{n=1}^{\infty}
\sum_{l=0}^{n-1}\lambda_{\rm s}^n
\frac{(-1)^{n-1-l}(n-1)!}{l!(n-1-l)!}
\partial_t\psi_{t,l}^{(\alpha,\beta)}|_{t=0}\\
&=&-\sum_{l=0}^{\infty}
\sum_{n=l+1}^{\infty}\lambda_{\rm s}^n
\frac{(-1)^{n-1-l}(n-1)!}{l!(n-1-l)!}
\partial_t\psi_{t,l}^{(\alpha,\beta)}|_{t=0}
=-\sum_{l=0}^{\infty}\lambda_S^{l+1}
\partial_t\psi_{t,l}^{(\alpha,\beta)}|_{t=0}\,,\nonumber
\end{eqnarray}
where the expansion parameter $\lambda_{\rm s}$ is redefined
as\footnote{
This fact is mentioned in Ref.~\citen{Erler_super}.
}
\begin{eqnarray}
 \lambda_{S}\equiv \frac{\lambda_{\rm s}}
{\lambda_{\rm s}+1}\,.
\label{eq:repara_lambda}
\end{eqnarray}
The last expression of the solution\footnote{
Such a solution for tachyon condensation is
also examined in Refs.~\citen{Okawa:2006sn, Erler:2006hw} and
\citen{Erler:2006ww}.
} can be formally summed as
\begin{eqnarray}
\Psi^{(\alpha,\beta)}&=&
Q_{\rm B}(\lambda_S P_{\alpha}*\hat{\Lambda}_0*P_{\beta})*\frac{1}{
1-\lambda_S P_{\alpha}*\hat{\Lambda}_0*P_{\beta}}\,,
\end{eqnarray}
which is the pure gauge form found in Ref.~\citen{Okawa}. 
Then, using
\begin{eqnarray}
\psi_{t,n}^{(\alpha,\beta)}&=&
P_{\alpha-\frac{1}{2}}*\psi_{t+n}|_{t=0}*
P_{\beta-\frac{1}{2}}
=(\alpha+\beta)
e^{\frac{\pi}{4}(\beta-\alpha)K_1}
(\alpha+\beta)^{\frac{D}{2}}\psi_{\frac{t}{\alpha+\beta}+n}
\,,
\end{eqnarray}
where $\psi_n$ is given in Ref.~\citen{Schnabl_tach} as
\begin{eqnarray}
 \psi_n&=&\frac{2}{\pi^2}U^{\dagger}_{n+2}U_{n+2}\!
\left[\hat{\cal B}\tilde c\biggl(\!-\frac{\pi n}{4}\biggr)
\tilde c\biggl(\frac{\pi n}{4}\biggr)\!+\frac{\pi}{2}
\biggl(\!\tilde c\biggl(\!-\frac{\pi n}{4}\biggr)+\tilde c\biggl(\frac{\pi
n}{4}\biggr)\!\biggr)
\right]\!|0\rangle,
\end{eqnarray}
we obtain (\ref{eq:rs_derivation}).
Because $\Psi^{(1/2,1/2)}$ coincides with the solution 
constructed in Ref.~\citen{Schnabl_tach} 
in this case [\,i.e.,
$\lambda_{\rm m}=0$ for $\hat{\psi}$ (\ref{eq:psi_sm})],
$\Psi^{(\alpha,\beta)}$ satisfies the generalized Schnabl gauge
condition ${\cal B}^{(\alpha,\beta)}\Psi^{(\alpha,\beta)}=0$
and should reproduce the D25-brane tension
for $\lambda_S=1$ $(\leftrightarrow \lambda_{\rm
s}=\infty)$,
\begin{eqnarray}
S[\Psi^{(\alpha,\beta)}]/V_{26}&=&\left\{
\begin{array}[tb]{cc}
\frac{1}{2\pi^2 g^2}&(\lambda_S=1)\\
 0& (|\lambda_S|<1)
\end{array}
\right.,\\
S[\Psi]&=&-\frac{1}{g^2}\left(
\frac{1}{2}\langle\Psi,Q_{\rm B}\Psi\rangle
+\frac{1}{3}\langle\Psi,\Psi*\Psi\rangle
\right)\,,
\label{eq:action_bos}
\end{eqnarray}
as evaluated in Refs.~\citen{Schnabl_tach, Okawa} and \citen{FK},
if we regularize it as
\begin{eqnarray}
\label{eq:reg_rs}
 \Psi^{(\alpha,\beta)}|_{\lambda_S=1}
&=&\lim_{N\to \infty}\left(
\frac{1}{\alpha+\beta}
\psi_{t=0,N}^{(\alpha,\beta)}
-\sum_{n=0}^N\partial_t\psi_{t,n}^{(\alpha,\beta)}|_{t=0}\right).
\end{eqnarray}
The first term can be added because $\psi_{t=0,N}^{(\alpha,\beta)}\sim
{\cal O}(((\alpha+\beta)N)^{-3})$ for $N\to \infty$ in the sense of
the $L_0$-level truncation.
With the above regularization, we find that the new BRST operator 
$Q_{\rm B}'$ around $\Psi^{(\alpha,\beta)}|_{\lambda_S=1}$
satisfies
\begin{eqnarray}
\label{eq:ES-gen}
Q_{\rm B}'A^{(\alpha+\beta)}\equiv Q_{\rm B}A^{(\alpha+\beta)}+
\Psi^{(\alpha,\beta)}|_{\lambda_S=1}*A^{(\alpha+\beta)}+A^{(\alpha+\beta)}*
\Psi^{(\alpha,\beta)}|_{\lambda_S=1}&=&I\,,~~~~
\end{eqnarray}
which implies a vanishing cohomology for $Q_{\rm B}'$,
at least formally in the sense of Ref.~\citen{ES}.

For each contribution to $\Psi^{(\alpha,\beta)}$
of a given power of $\lambda_S$, we have the relation 
\begin{eqnarray}
&&\partial_t\psi^{(\alpha,\beta)}_{t,n}|_{t=0}
-\partial_t\psi^{(\alpha,\beta)}_{t,n=0}|_{t=0}
=\frac{{\cal B}^{(\alpha,\beta)}}{{\cal L}^{(\alpha,\beta)}}
\sum_{m=0}^{n-1}
\partial_t\psi^{(\alpha,\beta)}_{t,m}|_{t=0}
*\partial_t\psi^{(\alpha,\beta)}_{t,n-1-m}|_{t=0}\,.~~~~(n\ge 1)
\nonumber\\
\label{eq:perturb_s}
\end{eqnarray}
Then, from (\ref{eq:rel_psis}), this implies 
\begin{eqnarray}
 \psi_{{\rm s},
k+1}&=&-\frac{{\cal B}^{(\alpha,\beta)}}{{\cal L}^{(\alpha,\beta)}}
\sum_{l=1}^{k}\psi_{{\rm s},l}*\psi_{{\rm s},k-l+1}\,
~~~~~(k\ge 1)
\end{eqnarray}
for each contribution to $\Psi^{(\alpha,\beta)}$
of a given power of $\lambda_{\rm s}$,
which has the same form as (\ref{eq:solok+1}).
The extra BRST exact term,
$-\partial_t\psi^{(\alpha,\beta)}_{t,n=0}|_{t=0}=\psi_{{\rm
s},1}=Q_{\rm B}(P_{\alpha}*\hat{\Lambda}_0*P_{\beta})$, 
in (\ref{eq:perturb_s}) induces the reparametrization
of the parameter (\ref{eq:repara_lambda})
when one solves the equation of motion (\ref{eq:EOM_bos}) perturbatively.
This is a simple example of the ambiguities discussed in
Ref.~\citen{FKP}.

\section{Solutions in superstring field theory
\label{sec:super}}

In the case of superstrings, we wish to solve
the equation of motion (\ref{eq:EOM_super})
in the large Hilbert space in the RNS formalism.
In particular, we use the $(\phi,\xi,\eta)$-system instead of
$(\beta,\gamma)$ for the superghost sector \cite{FMS}.
As in the bosonic case, we use a
$Q_{\rm B}$-invariant and $\eta_0$-invariant commutative monoid
$\{P_{\alpha}\}_{\alpha\ge 0}$, 
which has ghost number zero and picture number zero,
\begin{eqnarray}
&&Q_{\rm B}P_{\alpha}=0\,,~~~\eta_0P_{\alpha}=0\,,~~~
P_{\alpha}*P_{\beta}=P_{\alpha+\beta}\,,~~~P_{\alpha=0}=I,
\label{eq:comm_monoid_s}
\end{eqnarray}
and a counterpart to $A^{(\gamma)}$ given by
\begin{eqnarray}
&&\eta_0Q_{\rm B}\hat A^{(\gamma)}=I-P_{\gamma}\,.
\label{eq:hatAgamma}
\end{eqnarray}
Using these $P_{\alpha}$ and $\hat A^{(\gamma)}$ and a string
field $\hat{\phi}$ with ghost number zero and picture number zero
satisfying
\begin{eqnarray}
 \eta_0Q_{\rm B}\hat{\phi}=0\,,~~~~
\hat{\phi}*\hat{\phi}=0\,,~~~~
\hat{\phi}*\eta_0\hat{\phi}=0\,,~~~~
\hat{\phi}*Q_{\rm B}\hat{\phi}=0\,,
\label{eq:hatphi_super}
\end{eqnarray}
we can construct four types of solutions to the equation of motion:
\begin{eqnarray}
\Phi^{(\alpha,\beta)}_{(1)}&=&\log(1+P_{\alpha}*f_{(1)}*P_{\beta}),
~~~~~f_{(1)}=\frac{1}{1-\eta_0\hat{\phi}*Q_{\rm
B}\hat{A}^{(\alpha+\beta)}}*\hat{\phi},
\label{eq:Phi_1}
\\
\Phi^{(\alpha,\beta)}_{(2)}
&=&\log(1+P_{\alpha}*f_{(2)}*P_{\beta}),
~~~~~f_{(2)}=\hat{\phi}*\frac{1}{1-\eta_0\hat{A}^{(\alpha+\beta)}*Q_{\rm
B}\hat{\phi}},
\label{eq:Phi_2}\\
\Phi^{(\alpha,\beta)}_{(3)}&=&-\log(1-P_{\alpha}*f_{(3)}*P_{\beta}),
~~~~~f_{(3)}=\frac{1}{1
-Q_{\rm B}\hat{\phi}*\eta_0\hat{A}^{(\alpha+\beta)}}*\hat{\phi},
\label{eq:Phi_3}\\
\Phi^{(\alpha,\beta)}_{(4)}&=&-\log(1-P_{\alpha}*f_{(4)}*P_{\beta}),
~~~~~f_{(4)}=\hat{\phi}*\frac{1}{1
-Q_{\rm B}\hat{A}^{(\alpha+\beta)}*\eta_0\hat{\phi}}\,.
\label{eq:Phi_4}
\end{eqnarray}
In fact, from the derivation property of $Q_{\rm B}$ and $\eta_0$,
we obtain
\begin{eqnarray}
e^{-\Phi^{(\alpha,\beta)}_{(1)}}Q_{\rm
 B}e^{\Phi^{(\alpha,\beta)}_{(1)}}&=&P_{\alpha}*
\frac{1}{1-\eta_0(\hat{\phi}*Q_{\rm B}\hat{A}^{(\alpha+\beta)})}
*Q_{\rm B}\hat{\phi}*P_{\beta},\\
e^{-\Phi^{(\alpha,\beta)}_{(2)}}Q_{\rm
 B}e^{\Phi^{(\alpha,\beta)}_{(2)}}&=&P_{\alpha}*
Q_{\rm B}\hat{\phi}*
\frac{1}{1-\eta_0(\hat{A}^{(\alpha+\beta)}*Q_{\rm B}\hat{\phi})}
*P_{\beta},\\
e^{-\Phi^{(\alpha,\beta)}_{(3)}}Q_{\rm
 B}e^{\Phi^{(\alpha,\beta)}_{(3)}}
&=&P_{\alpha}*
\frac{1}{1+\eta_0(Q_{\rm B}\hat{\phi}*\hat{A}^{(\alpha+\beta)})}
*Q_{\rm B}\hat{\phi}*P_{\beta},\\
e^{-\Phi^{(\alpha,\beta)}_{(4)}}Q_{\rm
 B}e^{\Phi^{(\alpha,\beta)}_{(4)}}
&=&P_{\alpha}*
Q_{\rm B}\hat{\phi}*
\frac{1}{1+\eta_0(Q_{\rm B}\hat{A}^{(\alpha+\beta)}*\hat{\phi})}
*P_{\beta}\,.
\end{eqnarray}
These relations imply that the quantities
$\Phi^{(\alpha,\beta)}_{(i)}$ satisfy the following
equation of motion:
\begin{eqnarray}
 \eta_0\left(e^{-\Phi^{(\alpha,\beta)}_{(i)}}Q_{\rm B}
e^{\Phi^{(\alpha,\beta)}_{(i)}}\right)=0\,.~~~~(i=1,2,3,4)
\end{eqnarray}
Noting that the gauge transformation is given by
$e^{\Phi}\mapsto U*e^{\Phi}*V$ with $Q_{\rm B}U=0$ and $\eta_0V=0$,
the solutions given in 
 (\ref{eq:Phi_2})-(\ref{eq:Phi_3}) and 
(\ref{eq:Phi_1})-(\ref{eq:Phi_4})
are gauge equivalent because they are related as
\begin{eqnarray}
&&e^{\Phi^{(\alpha,\beta)}_{(2)}}
=U_{23}^{(\alpha,\beta)}* e^{\Phi^{(\alpha,\beta)}_{(3)}},~~
~~~~
e^{\Phi^{(\alpha,\beta)}_{(1)}}
= e^{\Phi^{(\alpha,\beta)}_{(4)}}*V_{41}^{(\alpha,\beta)}\,,
\label{eq:gauge_equiv1234}
\end{eqnarray}
where the gauge parameters are explicitly obtained as
\begin{eqnarray}
U_{23}^{(\alpha,\beta)}&\equiv&e^{\Phi^{(\alpha,\beta)}_{(2)}}
e^{-\Phi^{(\alpha,\beta)}_{(3)}}\nonumber\\
&=&I+P_{\alpha}*\biggl(
\hat{\phi}*\frac{1}{1-\eta_0\hat A^{(\alpha+\beta)}*Q_{\rm B}\hat{\phi}}
-\frac{1}{1-Q_{\rm B}\hat{\phi}*\eta_0\hat A^{(\alpha+\beta)}}*\hat{\phi}
\nonumber\\
&&-\hat{\phi}*\frac{1}{1
-\eta_0\hat A^{(\alpha+\beta)}*Q_{\rm B}\hat{\phi}}*P_{\alpha+\beta}
*\frac{1}{1-Q_{\rm B}\hat{\phi}*\eta_0\hat A^{(\alpha+\beta)}}*\hat{\phi}
\biggr)*P_{\beta}
\nonumber\\
&=&I-Q_{\rm B}\left(P_{\alpha}*
\hat{\phi}*\frac{1}{1
-\eta_0\hat A^{(\alpha+\beta)}*
Q_{\rm B}\hat{\phi}}*\eta_0\hat{A}^{(\alpha+\beta)}
*\hat{\phi}*P_{\beta}
\right),
\label{eq:U23}
\\
V_{41}^{(\alpha,\beta)}
&\equiv&e^{-\Phi^{(\alpha,\beta)}_{(4)}}
e^{\Phi^{(\alpha,\beta)}_{(1)}}\nonumber\\
&=&I+P_{\alpha}*\biggl(
-\hat{\phi}*\frac{1}{1-Q_{\rm B}\hat A^{(\alpha+\beta)}*\eta_0\hat{\phi}}
+\frac{1}{1-\eta_0\hat{\phi}*Q_{\rm
B}\hat A^{(\alpha+\beta)}}*\hat{\phi}\nonumber\\
&&-\hat{\phi}*\frac{1}{1-Q_{\rm B}\hat A^{(\alpha+\beta)}*\eta_0\hat{\phi}}
*P_{\alpha+\beta}*\frac{1}{1-\eta_0\hat{\phi}*Q_{\rm
B}\hat{\phi}}*\hat A^{(\alpha+\beta)}*\hat{\phi}
\biggr)*P_{\beta}
\nonumber\\
&=&I+\eta_0\left(P_{\alpha}*\hat{\phi}*
\frac{1}{1-Q_{\rm B}\hat{A}^{(\alpha+\beta)}*\eta_0\hat{\phi}}*Q_{\rm B}
\hat{A}^{(\alpha+\beta)}*\hat{\phi}*P_{\beta}\right).
\label{eq:V41}
\end{eqnarray}
Note that $\Phi^{(\alpha,\beta)}_{(i)}$ ($i=1,2,3,4$) all reduce to
$\hat{\phi}$ for $\alpha=\beta=0$, and
$\hat{\phi}$ given in (\ref{eq:hatphi_super}) is also a solution to 
the equation of motion (\ref{eq:EOM_super}).

As an example of $P_{\alpha}$, we consider
the family of wedge states (\ref{eq:wedgestates})
by replacing the total Virasoro generators with those of 
superstrings. 
Corresponding to this, we can construct $\hat A^{(\gamma)}$
(\ref{eq:hatAgamma}) as
\begin{eqnarray}
 \hat{A}^{(\gamma)}
&=&\int_0^{\gamma}d\alpha\log\left(\frac{\alpha}{\gamma}\right)
\left(\frac{\pi}{2}J_1^{--L}
+\alpha\frac{\pi^2}{4}\tilde G_1^{-L}B_1^L\right)P_{\alpha}\,,
\label{eq:hatA_wedge}
\end{eqnarray}
which satisfies
\begin{eqnarray}
\label{eq:hatAgamma_BG}
&&\eta_0\hat{A}^{(\gamma)}
=-\frac{\pi}{2}\int_0^{\gamma}\!d\alpha B_1^LP_{\alpha}\,,~~~~
 Q_{\rm B}\hat{A}^{(\gamma)}=-\frac{\pi}{2}\int_0^{\gamma}\!d\alpha 
\tilde G_1^{-L}P_{\alpha}\,,
\end{eqnarray}
and $\eta_0Q_{\rm B}\hat{A}^{(\gamma)}=I-P_{\gamma}$ (\ref{eq:hatAgamma}).
Here, $J_1^{--L}$ and $\tilde G_1^{-L}$
are defined in the same way as $B_1^L$, from $J^{--}(z)\equiv \xi b(z)$ 
and $\tilde G^{-}(z)\equiv [Q_{\rm B},J^{--}(z)]$. They are primary
fields of dimension 2 and are generators of the twisted $N=4$
superconformal algebra \cite{Berkovits:2001nr}. In terms of the mode
expansion, as in (\ref{eq:B0_B0d}), we have
\begin{eqnarray}
&&{\cal J}^{--}_0=\oint\frac{dw}{2\pi i}(1+w^2)(\arctan w) 
J^{--}(w)
=J^{--}_0
+\sum_{k=1}^{\infty}\frac{2(-1)^{k+1}}{4k^2-1}J^{--}_{2k}\,,\\
&&{\rm bpz}({\cal J}^{--}_0)=J^{--}_0
+\sum_{k=1}^{\infty}\frac{2(-1)^{k+1}}{4k^2-1}J^{--}_{-2k}\,,
~~~\hat{\cal J}^{--}\equiv {\cal J}^{--}_0+{\rm bpz}({\cal
J}^{--}_0)
\,,\\
&&J_1^{--L}=\int_{C_L} \frac{dz}{2\pi i}(1+z^2)J^{--}=\frac{1}{2}
(J_1^{--}+J^{--}_{-1})+\frac{1}{\pi}\hat{\cal J}^{--}\,,\\
&&\tilde{\cal G}^{-}_0=[Q_{\rm B},{\cal J}_0^{--}]
=\tilde G^{-}_0
+\sum_{k=1}^{\infty}\frac{2(-1)^{k+1}}{4k^2-1}\tilde G^{-}_{2k}\,,\\
&&{\rm bpz}(\tilde{\cal G}_0^{-})=
{\tilde G}^{-}_0
+\sum_{k=1}^{\infty}\frac{2(-1)^{k+1}}{4k^2-1}{\tilde G}^{-}_{-2k}\,,
~~~{\hat{\tilde {\cal G}}^-}\equiv 
\tilde{\cal G}^{-}_0+{\rm bpz}(\tilde{\cal G}^{-}_0)\,,\\
&&\tilde G_1^{-L}=[Q_{\rm B},J_1^{--L}]=\frac{1}{2}
(\tilde G_1^{-}+\tilde G^{-}_{-1})+\frac{1}{\pi}{\hat{\tilde {\cal
G}}^-}
\,,
\end{eqnarray}
and they satisfy the following (anti-)commutation relations 
obtained from their OPEs:
\begin{eqnarray}
&&[\eta_0,J_1^{--L}]=B_1^L,~~~\{\eta_0,\tilde G_1^{-L}\}=-K_1^L\,,
~~~\{Q_{\rm B},\tilde G_1^{-L}\}=0\,,\\
&&[J_1^{--L},\tilde G_1^{-L}]=0\,,~~~[K_1^{L},\tilde G_1^{-L}]=0\,,
~~~[K_1^L,J_1^{--L}]=0\,,
~~~\{B_1^L,\tilde G_1^{-L}\}=0\,.~~~~
\end{eqnarray}
We can derive a solution to  (\ref{eq:hatphi_super}) of the form
\begin{eqnarray}
 \hat{\phi}&=&\zeta_aU_1^{\dagger}U_1c\xi e^{-\phi}\psi^a(0)|0\rangle\,,
\label{eq:hatphi_1}
\end{eqnarray}
for example.
Here, $\psi^a$ represents the fermionic component of dimension $1/2$
of a supercurrent ${\bf
J}^a(z,\theta)=\psi^a(z)+\theta J^a(z)$
in the matter sector, and we assume that the OPEs are given by
\begin{eqnarray}
\psi^a(y)\psi^b(z)&\sim&(y-z)^{-1}\Omega^{ab}\,,
\label{eq:super_c_1}\\
J^a(y)\psi^b(z)&\sim&(y-z)^{-1}if^{ab}_{~~c}\psi^c(z)\,,
\label{eq:super_c_2}\\
J^a(y)J^b(z)&\sim&(y-z)^{-2}\Omega^{ab}
+(y-z)^{-1}if^{ab}_{~~c}J^c(z)\,,
\label{eq:super_c_3}
\end{eqnarray}
where $f^{ab}_{~~c}$  is the structure constant of the associated 
Lie algebra.
The quantities $\zeta_a$ are Grassmann even constants and satisfy
\begin{eqnarray}
 \zeta_a\zeta_b\Omega^{ab}=0\,,
\end{eqnarray}
which implies that $\zeta_a{\bf J}^a(z,\theta)$ is non-singular.
More concretely, we can take ${\bf J}^{\mu}(z,\theta)=\psi^{\mu}(z)+\theta
i\partial X^{\mu}(z)$ on a flat background, where $\zeta_{\mu}$ is directed
in a light-cone direction, so that
$\zeta_{\mu}\zeta_{\nu}\eta^{\mu\nu}=0$ holds.

The string field
$\hat A^{(\gamma)}$ given in (\ref{eq:hatA_wedge}) appears naturally
when we attempt to solve the equation of motion perturbatively.
More precisely, by substituting $\Phi=\sum_{n\ge 1}\lambda^n\Phi_n$ into
the equation of motion $\eta_0(e^{-\Phi}Q_{\rm B}e^{\Phi})=0$,
the first equation for  ${\cal O}(\lambda^1)$ is $\eta_0Q_{\rm
B}\Phi_1=0$, and the second one for ${\cal O}(\lambda^2)$ is
$\eta_0Q_{\rm B}\Phi_2=\frac{1}{2}(\eta_0\Phi_1*Q_{\rm B}\Phi_1+Q_{\rm
B}\Phi_1*\eta_0\Phi_1)$. In the case
$\Phi_1=P_{1/2}*\hat{\phi}*P_{1/2}$ with (\ref{eq:hatphi_1}),
the second-order term is computed as
\begin{eqnarray}
  \Phi_2&=&\frac{1}{2}\frac{\tilde{\cal G}^-_0}{{\cal L}_0}
\frac{{\cal B}_0}{ {\cal L}_0}
(\eta_0\Phi_1*Q_{\rm B}\Phi_1+Q_{\rm B}\Phi_1*\eta_0\Phi_1)
\nonumber\\
&=&-\frac{1}{2}P_{1/2}*
(\eta_0\hat\phi*\hat{A}^{(1)}*Q_{\rm B}\hat\phi+Q_{\rm
B}\hat\phi*\hat{A}^{(1)}
*\eta_0\hat\phi)*P_{1/2}\,.
\label{eq:Phi2sol}
\end{eqnarray}
In the first line, we have used the relation
$\frac{\tilde{\cal G}^-_0}{{\cal L}_0}
\frac{{\cal B}_0}{{\cal L}_0}
=\tilde{\cal G}^-_0\!\int_0^{\infty}
dT_1 e^{-T_1{\cal L}_0}{\cal B}_0\!\int_0^{\infty}
dT_2 e^{-T_2{\cal L}_0}$,
which is analogous to $\frac{\tilde{G}^-_0}{L_0}
\frac{b_0}{L_0}$ in Ref.~\citen{Michishita:2005se},
and explicitly carried out the computation in the sliver frame.
The result is identical to the last expression in (\ref{eq:Phi2sol}).
The above second-order 
term, $\Phi_2$, is equivalent to the ${\cal O}(\hat\phi^2)$
terms of (\ref{eq:Phi_1})--(\ref{eq:Phi_4}) with $\alpha=\beta=1/2$,
up to $\eta_0$-exact or $Q_{\rm B}$-exact terms.\\

We next comment on the reality condition for the string field $\Phi$.
In order to ensure the Berkovits' WZW-type superstring field action
in the NS sector,
\begin{eqnarray}
 S_{\rm NS}[\Phi]&=&-\frac{1}{g^2}\int_0^1dt \langle\!\langle
(\eta_0\Phi)(e^{-t\Phi}Q_{\rm B}e^{t\Phi})\rangle\!\rangle,
\label{eq:action_super}
\end{eqnarray}
to be real, we should impose the reality condition on the string field
$\Phi$. It is given by ${\rm bpz}^{-1}\!\circ{\rm hc}(\Phi)=-\Phi$.
Note that even if ${\rm bpz}^{-1}\!\circ{\rm hc}(\hat\phi)=-\hat\phi$,
the relation ${\rm bpz}^{-1}\!\circ{\rm
hc}(\Phi^{(\alpha,\beta)}_{(i)})=-\Phi^{(\alpha,\beta)}_{(i)}$
for (\ref{eq:Phi_1})--(\ref{eq:Phi_4})
is not satisfied in general.
However, we can construct two solutions, 
$\Phi^{(\alpha,\alpha)}_{(23)}$ and $\Phi^{(\alpha,\alpha)}_{(41)}$,
which satisfy the reality condition:
\begin{eqnarray}
&&e^{\Phi^{(\alpha,\alpha)}_{(23)}}
\equiv \sqrt{U_{23}^{(\alpha,\alpha)}}\,
* e^{\Phi^{(\alpha,\alpha)}_{(3)}},
~~~e^{\Phi^{(\alpha,\alpha)}_{(41)}}\equiv
 e^{\Phi^{(\alpha,\alpha)}_{(4)}}*\sqrt{V_{41}^{(\alpha,\alpha)}}\,,
\end{eqnarray}
with the gauge parameters (\ref{eq:U23}) and (\ref{eq:V41}).
(See also Ref.~\citen{Erler_super},v2.)
The square root is defined by an infinite series with respect to the star
product. Here we employ the conventions
\begin{eqnarray}
&&{\rm bpz}^{-1}\!\circ{\rm hc}(\hat{\phi})=-\hat{\phi}\,,~~~
{\rm bpz}^{-1}\!\circ{\rm hc}(\Phi_1*\Phi_2)=
{\rm bpz}^{-1}\!\circ{\rm hc}(\Phi_2)*
{\rm bpz}^{-1}\!\circ{\rm hc}(\Phi_1)\,,\nonumber\\
&&
{\rm bpz}^{-1}\!\circ{\rm hc}(\eta_0\Phi)=(-1)^{|\Phi|}\eta_0
{\rm bpz}^{-1}\!\circ{\rm hc}(\Phi),\nonumber\\
&&{\rm bpz}^{-1}\!\circ{\rm hc}(Q_{\rm B}\Phi)
=-(-1)^{|\Phi|}
Q_{\rm B}{\rm bpz}^{-1}\!\circ{\rm hc}(\Phi),\nonumber\\
&&{\rm bpz}^{-1}\!\circ{\rm hc}(\eta_0\hat{A}^{(\alpha+\beta)})
=\eta_0\hat{A}^{(\alpha+\beta)}\,,
~~~{\rm bpz}^{-1}\!\circ{\rm hc}(Q_{\rm B}\hat{A}^{(\alpha+\beta)})
=-Q_{\rm B}\hat{A}^{(\alpha+\beta)}\,,~~
\end{eqnarray}
and note the relations ${\rm bpz}^{-1}\!\circ{\rm hc}(f_{(2)})=-f_{(3)}$ and
 ${\rm bpz}^{-1}\!\circ{\rm hc}(f_{(1)})=-f_{(4)}$.

Alternatively, using the method in Ref.~\citen{Okawa_real}, we have
a solution $\Phi_{\lambda}$ of the path-ordered form:
\begin{eqnarray}
e^{\Phi_{\lambda}}&=&{\rm
 P}\exp\int_0^{\lambda}d\lambda'G(\lambda')=I+\!
\int_0^{\lambda}\!d\lambda'G(\lambda')+\!
\int_0^{\lambda}\!d\lambda_1\!
\int_0^{\lambda_1}\!d\lambda_2
G(\lambda_2)*G(\lambda_1)+\cdots,\nonumber\\
\label{eq:sol_okawa_real}\\
G(\lambda)&=&P_{\alpha}*\frac{1}{1-\lambda Q_{\rm
 B}\hat{\phi}*\eta_0\hat{A}^{(\alpha+\beta)}}
*\hat{\phi}*\frac{1}{1-\lambda\eta_0\hat{A}^{(\alpha+\beta)}*Q_{\rm
B}\hat{\phi}}*P_{\beta}\,.
\end{eqnarray}
In fact, $G(\lambda)$ is a solution of the relation
\begin{eqnarray}
\label{eq:Gdiff}
&&Q_{\rm
 B}G(\lambda)+[\Psi_{\lambda},G(\lambda)]
=\frac{d}{d\lambda}\Psi_{\lambda},
\end{eqnarray}
with
\begin{eqnarray}
&&\Psi_{\lambda}\equiv P_{\alpha}*\frac{1}{1-\lambda Q_{\rm
 B}\hat{\phi}*\eta_0\hat{A}^{(\alpha+\beta)}}*\lambda Q_{\rm
 B}\hat{\phi}*P_{\beta},
\end{eqnarray}
where the above $\Psi_{\lambda}$ satisfies 
\begin{eqnarray}
&&Q_{\rm B}\Psi_{\lambda}+\Psi_{\lambda}*\Psi_{\lambda}=0\,,~~~
\eta_0\Psi_{\lambda}=0\,.\label{eq:EOMPsi}
\end{eqnarray}
In addition, the above $e^{\Phi_{\lambda}}$ (\ref{eq:sol_okawa_real})
satisfies
\begin{eqnarray}
Q_{\rm
 B}G(\lambda)+[(e^{-\Phi_{\lambda}}Q_{\rm B}e^{\Phi_{\lambda}}) ,G(\lambda)]=
\frac{d}{d\lambda}(e^{-\Phi_{\lambda}}Q_{\rm B}e^{\Phi_{\lambda}}),
~~~(e^{-\Phi_{\lambda}}Q_{\rm B}e^{\Phi_{\lambda}}) |_{\lambda=0}=0.
~~~~~~~
\end{eqnarray}
Comparing them to (\ref{eq:Gdiff}), we conclude the relation
$\Psi_{\lambda}=e^{-\Phi_{\lambda}}Q_{\rm B}e^{\Phi_{\lambda}}$.
Therefore, the second equation in
(\ref{eq:EOMPsi}) implies that $\Phi_{\lambda}$ is a solution to
the equation of motion (\ref{eq:EOM_super}).
Because we have ${\rm bpz}^{-1}\!\circ {\rm hc}(G(\lambda))
=-G(\lambda)$ for real $\lambda$ and $\alpha=\beta$, this
$\Phi_{\lambda}$ with $\alpha=\beta$ satisfies the reality condition,
as in Ref.~\citen{Okawa_real}.

We can similarly obtain another solution $\tilde \Phi_{\lambda}$ of
path-ordered form:
\begin{eqnarray}
e^{-\tilde \Phi_{\lambda}}&=&{\rm
 P}\exp\int_0^{\lambda}d\lambda'\tilde G(\lambda')
\,,~~~~~
\label{eq:sol_okawa_real2}\\
\tilde G(\lambda)&=&-P_{\alpha}*\frac{1}{1-\lambda \eta_0
\hat{\phi}*Q_{\rm B}\hat{A}^{(\alpha+\beta)}}
*\hat{\phi}*\frac{1}{1-\lambda Q_{\rm
B}\hat{A}^{(\alpha+\beta)}*\eta_0\hat{\phi}}*P_{\beta}\,.
\end{eqnarray}
In this case, $\tilde G(\lambda)$ is a solution of
\begin{eqnarray}
\label{eq:Gdiff2}
&&\eta_0\tilde G(\lambda)
+[\tilde\Psi_{\lambda},\tilde G(\lambda)]=\frac{d}{d\lambda}
\tilde\Psi_{\lambda},~~~\\
&&\tilde\Psi_{\lambda}\equiv -P_{\alpha}*\frac{1}{1-\lambda
\eta_0\hat{\phi}*Q_{\rm B}\hat{A}^{(\alpha+\beta)}}*\lambda \eta_0
\hat{\phi}*P_{\beta},~~~
\end{eqnarray}
where $\tilde\Psi_{\lambda}$ satisfies 
\begin{eqnarray}
&&\eta_0\tilde\Psi_{\lambda}+\tilde\Psi_{\lambda}*\tilde\Psi_{\lambda}=0\,,
~~~~~Q_{\rm B}\tilde\Psi_{\lambda}=0\,.\label{eq:EOMPsi2}
\end{eqnarray}
In addition, the above $e^{-\tilde\Phi_{\lambda}}$
(\ref{eq:sol_okawa_real2})
satisfies
\begin{eqnarray}
\eta_0\tilde G(\lambda)
+[(e^{\tilde\Phi_{\lambda}}\eta_0 e^{-\tilde\Phi_{\lambda}}),\tilde
G(\lambda)]=
\frac{d}{d\lambda}(e^{\tilde\Phi_{\lambda}}\eta_0 e^{-\tilde\Phi_{\lambda}})\,,
~~~(e^{\tilde\Phi_{\lambda}}\eta_0e^{-\tilde\Phi_{\lambda}})|_{\lambda=0}=0\,.
~~~
\end{eqnarray}
Comparing them to (\ref{eq:Gdiff2}), we conclude the relation
$\tilde\Psi_{\lambda}=e^{\tilde\Phi_{\lambda}}\eta_0
e^{-\tilde\Phi_{\lambda}}$.
Therefore, the second equation in 
(\ref{eq:EOMPsi2}) implies that $\tilde\Phi_{\lambda}$ is a solution to
the equation of motion, owing to the relation
\begin{eqnarray}
 \eta_0(e^{-{\tilde \Phi}_{\lambda}}Q_{\rm B}e^{\tilde \Phi_{\lambda}})
&=&e^{-{\tilde \Phi}_{\lambda}}(Q_{\rm B}
(e^{\tilde\Phi_{\lambda}}\eta_0e^{-\tilde\Phi_{\lambda}}))
e^{\tilde \Phi_{\lambda}}=0\,.
\end{eqnarray}
Because we have ${\rm bpz}^{-1}\!\circ {\rm hc}(\tilde G(\lambda))
=-\tilde G(\lambda)$ for real $\lambda$
and $\alpha=\beta$,
 this $\tilde\Phi_{\lambda}$ with $\alpha=\beta$ satisfies the reality
condition.

\section{Gauge transformations and field redefinitions
\label{sec:gauge_tr}}

\subsection{Finite gauge transformations
\label{sec:gauge}}

Let us consider gauge transformations among the obtained solutions
in bosonic and super string field theory.
We find that
$\Psi^{(\alpha,\beta)}$ given in (\ref{eq:b_sol_gen})
and $\Phi^{(\alpha,\beta)}_{(i)}$ given in
(\ref{eq:Phi_1})--(\ref{eq:Phi_4})
can be rewritten as particular finite gauge
transformations from simple solutions $\hat{\psi}$ and $\hat{\phi}$,
respectively,
in bosonic and super string field theory
{\it if we formally use $P_{\alpha}^{-1}$ 
or, equivalently, we formally extend a commutative monoid
$\{P_{\alpha}\}_{\alpha\ge 0}$ to an
Abelian group with respect to the star product}.
In the case of a bosonic string, we have
\begin{eqnarray}
\label{eq:gauge_bos}
&&\Psi^{(\alpha,\beta)}=V^{-1}*\hat{\psi}*V+V^{-1}*Q_{\rm B}V\,,\\
&&V=(I+\hat{\psi}*A^{(\alpha+\beta)})*P_{\alpha}^{-1},~~~
V^{-1}=P_{\alpha}*\frac{1}{1+\hat{\psi}*A^{(\alpha+\beta)}}\,.
\end{eqnarray}
Similarly, for a superstring, we obtain
\begin{eqnarray}
\label{eq:gauge_super3}
 e^{\Phi^{(\alpha,\beta)}_{(3)}}&=&P_{\alpha}*\frac{1}{1-Q_{\rm
 B}(\hat{\phi}*\eta_0\hat{A}^{(\alpha+\beta)})}
*e^{\hat{\phi}}*
(I+\eta_0(Q_{\rm
 B}\hat{\phi}*\hat{A}^{(\alpha+\beta)}))*P_{\alpha}^{-1},\\
\label{eq:gauge_super4}
e^{\Phi^{(\alpha,\beta)}_{(4)}}&=&P_{\beta}^{-1}*(I-Q_{\rm
 B}(\hat{A}^{(\alpha+\beta)}*\eta_0\hat{\phi}))
*e^{\hat{\phi}}*\frac{1}{1+\eta_0(Q_{\rm
 B}\hat{A}^{(\alpha+\beta)}*\hat{\phi})}*P_{\beta},
\end{eqnarray}
and $\Phi^{(\alpha,\beta)}_{(i)}$ for $i=1,2$ are related by
(\ref{eq:gauge_equiv1234}).
More precisely, if $\{P_{\alpha}\}_{\alpha\ge 0}$ can be extended 
 to an Abelian group, all the solutions $\Psi^{(\alpha,\beta)}$ are
 gauge equivalent
in bosonic string field theory, and all the solutions
$\Phi^{(\alpha,\beta)}_{(i)}$($i=1,2,3,4$) are gauge 
equivalent in superstring field theory.
In the case that $\{P_{\alpha}\}_{\alpha\ge 0}$ is the family of wedge
states, we can formally rewrite $P_{\alpha}$ as 
$\exp(-\frac{\pi}{2}\alpha K_1^L I)$
by using $(K_1^L\Phi_1)*\Phi_2=K_1^L(\Phi_1*\Phi_2)$ and the identity
property. Therefore, we can regard $\{P_{\alpha}\}_{\alpha\in
\mathbb{R}}$ with $P_{\alpha}^{-1}=P_{-\alpha}$ as an Abelian group.
However, 
$P_{\alpha}^{-1}=P_{-\alpha}$ ($\alpha>0$) might not be valid as a
wedge state, due to its ``negative angle.''
In any case, it seems that
our solutions $\Psi^{(\alpha,\beta)}$ and
$\Phi^{(\alpha,\beta)}_{(i)}$ are almost gauge equivalent to 
$\Psi^{(\alpha',\beta')}$ and $\Phi^{(\alpha',\beta')}_{(i)}$ 
with $(\alpha,\beta)\ne (\alpha',\beta')$, respectively, in this sense.

Alternatively, we can obtain finite gauge parameter string fields of
path-ordered forms in the case that $\{P_{\alpha}\}_{\alpha\ge 0}$ is
the family of wedge states, using the expressions given in
(\ref{eq:wedgestates}), (\ref{eq:Agamma_ex}) and
(\ref{eq:hatAgamma_BG}).

In the case of bosonic string field theory, as in Ref.~\citen{Ellwood:2007xr},
we find
\begin{eqnarray}
&&Q_{\rm
 B}G^{(\alpha,\beta)}(t)+[\Psi^{(t\alpha,t\beta)},G^{(\alpha,\beta)}(t)]
=\frac{d}{dt}\Psi^{(t\alpha,t\beta)},
~~~~~\Psi^{(t\alpha,t\beta)}|_{t=0}=\hat\psi\,,\\
&&G^{(\alpha,\beta)}(t)\equiv -\frac{\pi}{2}(\alpha B_1^L-\beta B_1^R)
\Psi^{(t\alpha,t\beta)}
\\
&&~~~~~~~~~~~\,=\frac{-\pi}{2}
\biggl(\alpha( B_1^L
P_{t\alpha})
*\frac{1}{1+\hat{\psi}*A^{(t\alpha+t\beta)}}*\hat{\psi}*P_{t\beta}
\nonumber\\
&&~~~~~~~~~~~~~~~~~~~~~\,+\beta
P_{t\alpha}
*\frac{1}{1+\hat{\psi}*A^{(t\alpha+t\beta)}}*\hat{\psi}
*B_1^RP_{t\beta}
\biggr).\nonumber
\end{eqnarray}
This implies the relation
\begin{eqnarray}
\Psi^{(\alpha,\beta)}&=&V^{(\alpha,\beta)-1}*\hat\psi*V^{(\alpha,\beta)}
+V^{(\alpha,\beta)-1}*Q_{\rm B}*V^{(\alpha,\beta)}\,,
\end{eqnarray}
with
\begin{eqnarray}
V^{(\alpha,\beta)}&=&{\rm
 P}\exp\int_0^1dtG^{(\alpha,\beta)}(t)\,,
\end{eqnarray}
where the path-order ${\rm P}$ is defined as in (\ref{eq:sol_okawa_real}),
so that it satisfies $\frac{d}{dt}{\rm P}e^{\int_0^tdu G(u)}=({\rm
P}e^{\int_0^tduG(u)})G(t)$. \sloppy
Therefore, all the solutions $\Psi^{(\alpha,\beta)}$
are gauge equivalent to the simplest one, $\hat{\psi}$.

In the case of superstring field theory, rewriting (\ref{eq:Phi_3}) as
\begin{eqnarray}
&&e^{\Phi^{(\alpha,\beta)}_{(3)}}
=I+P_{\alpha}*\frac{1}{1-Q_{\rm B}
(\hat{\phi}*\eta_0\hat{A}^{(\alpha+\beta)})}*\hat{\phi}*P_{\beta},
\end{eqnarray}
$\frac{d}{dt}e^{\Phi^{(t\alpha,t\beta)}_{(3)}}$ is computed as
\begin{eqnarray}
\frac{d}{dt}e^{\Phi^{(t\alpha,t\beta)}_{(3)}}&=&
\beta\frac{\pi}{2}K_1^R\!\left(e^{\Phi^{(t\alpha,t\beta)}_{(3)}}-I\right)
-\alpha\frac{\pi}{2}K_1^L\!\left(e^{\Phi^{(t\alpha,t\beta)}_{(3)}}
-I\right)\nonumber\\
&&-\frac{\pi}{2}(\alpha+\beta)P_{t\alpha}*
\frac{1}{1-Q_{\rm B}(\hat\phi*\eta_0\hat{A}^{(t\alpha+t\beta)})}
*Q_{\rm B}(\hat{\phi}*B_1^LP_{t(\alpha+\beta)})\nonumber\\
&&~~~~~~~~~~~~~~~~~~~*\frac{1}{1-Q_{\rm
 B}(\hat\phi*\eta_0\hat{A}^{(t\alpha+t\beta)})}*\hat{\phi}*P_{t\beta}
\nonumber\\
&=&\beta\frac{\pi}{2}K_1^R\!\left(e^{\Phi^{(t\alpha,t\beta)}_{(3)}}\!\!\!-I
\right)
-\alpha\frac{\pi}{2}K_1^L\!\left(e^{\Phi^{(t\alpha,t\beta)}_{(3)}}\!\!\!-I
\right)
\nonumber\\
&&+\frac{\pi}{2}(\alpha+\beta)\!
\left(Q_{\rm B}B_1^R\!\left(e^{\Phi^{(t\alpha,t\beta)}_{(3)}}\!\!\!-I
\right)
\right)\!*\!\left(e^{\Phi^{(t\alpha,t\beta)}_{(3)}}\!\!-I\right)\nonumber\\
&=&G_{1}^{(\alpha,\beta)}(t)*e^{\Phi^{(t\alpha,t\beta)}_{(3)}}
+e^{\Phi^{(t\alpha,t\beta)}_{(3)}}*G_{2}^{(\alpha,\beta)}(t)\,,
\end{eqnarray}
where
\begin{eqnarray}
G_{1}^{(\alpha,\beta)}(t)&\equiv& -\frac{\pi}{2}\alpha K_1^LI+
\frac{\pi}{2}(\alpha+\beta)
Q_{\rm B}B_1^R\!\left(e^{\Phi^{(t\alpha,t\beta)}_{(3)}}\!\!-I\right)\\
&=&-\frac{\pi}{2}\alpha K_1^LI+\frac{\pi}{2}(\alpha+\beta)
P_{t\alpha}*\frac{1}{1-Q_{\rm B}
(\hat{\phi}*\eta_0\hat{A}^{(t\alpha+t\beta)})}*Q_{\rm
B}(\hat{\phi}*B_1^RP_{t\beta})\,,\nonumber\\
G_{2}^{(\alpha,\beta)}(t)&\equiv& \frac{\pi}{2}\alpha K_1^LI+
\frac{\pi}{2}(\alpha+\beta)
B_1^R\!\left(e^{-\Phi^{(t\alpha,t\beta)}_{(3)}}Q_{\rm B}
e^{\Phi^{(t\alpha,t\beta)}_{(3)}}\right)
\\
&=&
\frac{\pi}{2}\alpha K_1^LI
-\frac{\pi}{2}(\alpha+\beta)
P_{t\alpha}*
\frac{1}{1+\eta_0(Q_{\rm B}\hat{\phi}*\hat{A}^{(t\alpha+t\beta)})}
*Q_{\rm B}\hat{\phi}*B_1^RP_{t\beta}\,.
\nonumber
\label{eq:G2}
\end{eqnarray}
The above $G_{1}^{(\alpha,\beta)}(t)$ and $G_{2}^{(\alpha,\beta)}(t)$ 
satisfy
\begin{eqnarray}
Q_{\rm B}G_{1}^{(\alpha,\beta)}(t)=0\,,~~~~~
 \eta_0G_{2}^{(\alpha,\beta)}(t)=0\,,
\end{eqnarray}
which are conditions for gauge parameters in superstring field theory.
Therefore, we find a finite gauge transformation
using the path-ordered form:
\begin{eqnarray}
\label{eq:gauge_3_path}
&&e^{\Phi^{(\alpha,\beta)}_{(3)}}=W_1*e^{\hat{\phi}}*W_2\,,
~~~~Q_{\rm B}W_1=0,~~~~~\eta_0W_2=0\,,\\
&&W_1\equiv{\rm
 P}'\exp\int_0^1dtG_1^{(\alpha,\beta)}(t)\,,~~~~
W_2\equiv{\rm
 P}\exp\int_0^1dtG_2^{(\alpha,\beta)}(t)\,.
\label{eq:W12}
\end{eqnarray}
Here, ${\rm P}'$ is the path order opposite to ${\rm P}$.
Thus, we have
$\frac{d}{dt}{\rm P}'e^{\int_0^tdu G(u)}=G(t)({\rm P}'e^{\int_0^tdu
G(u)})$, \sloppy
 or, explicitly,
\begin{eqnarray}
 {\rm P}'e^{\int_0^tdu G(u)}&=&
I+\!
\int_0^{\lambda}\!d\lambda'G(\lambda')+\!
\int_0^{\lambda}\!d\lambda_1\!
\int_0^{\lambda_1}\!d\lambda_2
G(\lambda_1)*G(\lambda_2)+\cdots.
\end{eqnarray}
Similarly, from the expression of
$\Phi^{(\alpha,\beta)}_{(1)}$ given in (\ref{eq:Phi_1}), we obtain
\begin{eqnarray}
e^{-\Phi^{(\alpha,\beta)}_{(1)}}
&=&I-P_{\alpha}*\frac{1}{1-\eta_0(\hat{\phi}*Q_{\rm
  B}\hat{A}^{(\alpha+\beta)})}*\hat{\phi}*P_{\beta}\,,
\end{eqnarray}
and this yields
\begin{eqnarray}
\frac{d}{dt}\,e^{-\Phi^{(t\alpha,t\beta)}_{(1)}}
&=&-G_{3}^{(\alpha,\beta)}(t)*e^{-\Phi^{(t\alpha,t\beta)}_{(1)}}
-e^{-\Phi^{(t\alpha,t\beta)}_{(1)}}*G_{4}^{(\alpha,\beta)}(t)\,,
\end{eqnarray}
where
\begin{eqnarray}
 G_{3}^{(\alpha,\beta)}(t)&\equiv&
\frac{\pi}{2}\alpha K_1^L I+\frac{\pi}{2}(\alpha+\beta)\eta_0
\tilde G_1^{-R}\!\left(e^{-\Phi^{(t\alpha,t\beta)}_{(1)}}-I\right)\\
&=&\frac{\pi}{2}\alpha K_1^LI
-\frac{\pi}{2}(\alpha+\beta)P_{t\alpha}*
\frac{1}{1-\eta_0(\hat{\phi}*Q_{\rm
  B}\hat{A}^{(t\alpha+t\beta)})}
*\eta_0(\hat{\phi}*\tilde G_1^{-R}P_{t\beta})\,,
\nonumber\\
 G_{4}^{(\alpha,\beta)}(t)&\equiv&-\frac{\pi}{2}\alpha K_1^LI
+\frac{\pi}{2}(\alpha+\beta)\tilde G_1^{-R}
\!\left(e^{\Phi^{(t\alpha,t\beta)}_{(1)}}
\eta_0e^{-\Phi^{(t\alpha,t\beta)}_{(1)}}\right)\\
&=&
-\frac{\pi}{2}\alpha K_1^LI
+\frac{\pi}{2}(\alpha+\beta)P_{t\alpha}*
\frac{1}{1+Q_{\rm B}(\eta_0\hat{\phi}*\hat{A}^{(t\alpha+t\beta)})}
*\eta_0\hat{\phi}*\tilde G_1^{-R}P_{t\beta}\,.
\nonumber
\end{eqnarray}
These quantities $G_{1}^{(\alpha,\beta)}(t)$ and
$G_{2}^{(\alpha,\beta)}(t)$ satisfy the relations
\begin{eqnarray}
\eta_0G_{3}^{(\alpha,\beta)}(t)=0\,,~~~~~
Q_{\rm B}G_{4}^{(\alpha,\beta)}(t)=0\,.
\end{eqnarray}
Then, we obtain a finite gauge transformation 
from  $\hat\phi$:
\begin{eqnarray}
\label{eq:gauge_1_path}
&&e^{\Phi^{(\alpha,\beta)}_{(1)}}=W_3*e^{\hat{\phi}}*W_4\,,
~~~~Q_{\rm B}W_3=0,~~~~~\eta_0W_4=0\,,\\
&&W_3\equiv{\rm
 P}'\exp\int_0^1dtG_4^{(\alpha,\beta)}(t)\,,~~~~
W_4\equiv{\rm
 P}\exp\int_0^1dtG_3^{(\alpha,\beta)}(t)\,.
\label{eq:W34}
\end{eqnarray}
The solutions $\Phi^{(\alpha,\beta)}_{(2)}$ in (\ref{eq:Phi_2}) and 
$\Phi^{(\alpha,\beta)}_{(4)}$ in (\ref{eq:Phi_4})
are also gauge equivalent to $\hat{\phi}$, by (\ref{eq:gauge_3_path}),
(\ref{eq:gauge_1_path}) and (\ref{eq:gauge_equiv1234}).
Therefore, our solutions
$\Phi^{(\alpha,\beta)}_{(i)}$ appearing in
(\ref{eq:Phi_1})--(\ref{eq:Phi_4}) are all gauge equivalent to 
the simplest one, $\hat\phi$, in superstring field theory
with the gauge parameters in the path-ordered forms (\ref{eq:W12})
and (\ref{eq:W34}).

Although our solutions 
$\Psi^{(\alpha,\beta)}$ and $\Phi^{(\alpha,\beta)}_{(i)}$
are all gauge equivalent to each other 
in bosonic and super string field theory, respectively,
in the above sense, we note that
the gauge parameter string fields might become singular.
For example, as observed in Ref.~\citen{Okawa}, 
Schnabl's solution for tachyon condensation \cite{Schnabl_tach}
can be regarded as a limit of a pure gauge form.
A similar situation is found in Ref.~\citen{TT_tach}.
Therefore, we must study the ``regularity'' of 
gauge parameter string fields more carefully 
in order to conclude gauge equivalence.

\subsection{Pure gauge forms and induced field redefinitions}

If $\hat{\psi}$ and $\hat{\phi}$ can be rewritten
in pure gauge forms, i.e.
\begin{eqnarray}
\label{eq:puregauge_b}
&&\hat{\psi}=e^{-\Lambda}Q_{\rm B}e^{\Lambda},\\
&&e^{\hat{\phi}}=e^{Q_{\rm B}\Lambda_1}e^{\eta_0\Lambda_2},
\label{eq:puregauge_s}
\end{eqnarray}
then $\Psi^{(\alpha,\beta)}$ and $\Phi^{(\alpha,\beta)}_{(i)}$
can also be rewritten in pure gauge forms
without path-ordered forms as
\begin{eqnarray}
\Psi^{(\alpha,\beta)}
&=&U^{(\alpha,\beta)-1} Q_{\rm B}U^{(\alpha,\beta)}\,,
\nonumber\\
U^{(\alpha,\beta)}&=&I
+P_{\alpha}*(e^{\Lambda}-I)*\frac{1}{1+A^{(\alpha+\beta)}
*\hat{\psi}}*P_{\beta}\,,
\label{eq:pure_gauge_b}
\end{eqnarray}
in bosonic string field theory, and as
\begin{eqnarray}
 e^{\Phi^{(\alpha,\beta)}_{(i)}}&=&U_{(i)}^{(\alpha,\beta)}
*V_{(i)}^{(\alpha,\beta)}\,,~~~
Q_{\rm
B}U_{(i)}^{(\alpha,\beta)}=0,~~~\eta_0V_{(i)}^{(\alpha,\beta)}=0\,,
\label{eq:pure_gauge_s}
\end{eqnarray}
in superstring field theory, where we have the following:
\begin{eqnarray}
V_{(3)}^{(\alpha,\beta)}&=&V_{(2)}^{(\alpha,\beta)}=I+P_{\alpha}
*(e^{\eta_0\Lambda_2}-I)*\frac{1}{
1-\eta_0\hat{A}^{(\alpha+\beta)}*Q_{\rm B}\hat{\phi}}*P_{\beta}\,,\\
U_{(3)}^{(\alpha,\beta)}&=&\left[
I+P_{\alpha}*\frac{1}{1-Q_{\rm B}
(\hat{\phi}*\eta_0\hat{A}^{(\alpha+\beta)})}*\hat{\phi}*P_{\beta}
\right]*V_{(3)}^{(\alpha,\beta)-1}\,,\\
U_{(4)}^{(\alpha,\beta)}&=&U_{(1)}^{(\alpha,\beta)}=
I+P_{\alpha}*\frac{1}{
1-\eta_0\hat{\phi}*Q_{\rm B}\hat{A}^{(\alpha+\beta)}}*
(e^{Q_{\rm B}\Lambda_1}-I)*P_{\beta}\,,\\
V_{(4)}^{(\alpha,\beta)}&=&U_{(4)}^{(\alpha,\beta)-1}*\left[
I+P_{\alpha}*\hat{\phi}*
\frac{1}{1+\eta_0(Q_{\rm B}\hat{A}^{(\alpha+\beta)}
*\hat{\phi})}*P_{\beta}
\right]\,,\\
V_{(1)}^{(\alpha,\beta)}&=&
V_{(4)}^{(\alpha,\beta)}*V_{41}^{(\alpha,\beta)}\,,\\
U_{(2)}^{(\alpha,\beta)}&=&U_{23}^{(\alpha,\beta)}*U_{(3)}^{(\alpha,\beta)}
\,.
\end{eqnarray}
Here, we note that
\begin{eqnarray}
 V_{(3)}^{(\alpha,\beta)-1}*Q_{\rm B}V_{(3)}^{(\alpha,\beta)}
&=&P_{\alpha}*Q_{\rm
 B}\hat{\phi}*
\frac{1}{1-\eta_0\hat{A}^{(\alpha+\beta)}*Q_{\rm B}\hat{\phi}}*P_{\beta}\,,\\
\eta_0U_{(4)}^{(\alpha,\beta)}*U_{(4)}^{(\alpha,\beta)-1}
&=&P_{\alpha}*\frac{1}{1
-\eta_0\hat{\phi}*Q_{\rm B}\hat{A}^{(\alpha+\beta)}}
*\eta_0\hat{\phi}*P_{\beta}\,,
\end{eqnarray}
are useful to check the conditions for gauge parameters.
In the expressions of
(\ref{eq:pure_gauge_b}) and (\ref{eq:pure_gauge_s}), 
we do not have to use $P_{\alpha}^{-1}$,
unlike in the case of (\ref{eq:gauge_bos}) and
(\ref{eq:gauge_super3})-(\ref{eq:gauge_super4}).
In this sense, they are well-defined if the
gauge parameter string fields,
i.e. $\Lambda$ in (\ref{eq:puregauge_b})
and $\Lambda_1$ and $\Lambda_2$ in (\ref{eq:puregauge_s}), 
are well-defined.\\

Around the solutions $\Psi^{(\alpha,\beta)}$
and $\Phi^{(\alpha,\beta)}_{(i)}$
in the forms of (\ref{eq:pure_gauge_b}) and (\ref{eq:pure_gauge_s}),
the action $S[\Psi]$ given in (\ref{eq:action_bos})
for a bosonic string field
and $S_{\rm NS}[\Phi]$  given in (\ref{eq:action_super})
for a superstring field in the NS sector can be re-expanded as\footnote{
See also Ref.~\citen{Kluson}.
}
\begin{eqnarray}
 S[\Psi^{(\alpha,\beta)}
+\Psi]&=& S[\Psi^{(\alpha,\beta)}]
+ S[U^{(\alpha,\beta)}*\Psi*U^{(\alpha,\beta)-1}]\,,\\
 S_{\rm NS}[\log(e^{\Phi^{(\alpha,\beta)}_{(i)}}e^{\Phi})]
&=& S_{\rm NS}[\Phi^{(\alpha,\beta)}_{(i)}]
+ S_{\rm NS}[V_{(i)}^{(\alpha,\beta)}*\Phi
*V_{(i)}^{(\alpha,\beta)-1}]\,.
\end{eqnarray}
The second terms induce the following field redefinitions:
\begin{eqnarray}
\label{eq:redef_b}
 U^{(\alpha,\beta)}*\Psi*U^{(\alpha,\beta)-1}
&=&\Psi+(P_{\alpha}*\Lambda*P_{\beta})*\Psi
-\Psi*(P_{\alpha}*\Lambda*P_{\beta})
+{\cal O}(\Lambda^2),~~~\\
\label{eq:redef_s}
 V_{(i)}^{(\alpha,\beta)}*\Phi*V_{(i)}^{(\alpha,\beta)-1}
&=&\Phi+(P_{\alpha}*\eta_0\Lambda_2*P_{\beta})
*\Phi-\Phi*(P_{\alpha}*\eta_0\Lambda_2*P_{\beta})\\
&&+{\cal O}(\Lambda_1^2,\Lambda_1\Lambda_2,\Lambda_2^2),~~~\nonumber
\end{eqnarray}
respectively. \\

In particular, for the solutions $\Psi^{(\alpha,\beta)}$
and  $\Phi^{(\alpha,\beta)}_{(i)}$ generated from
\begin{eqnarray}
\hat{\psi}&=&\zeta_{\mu}U_1^{\dagger} U_1c i\partial X^{\mu}(0)|0\rangle,\\
\hat{\phi}&=&\zeta_{\mu}U_1^{\dagger} U_1 c\xi e^{-\phi}
 \psi^{\mu}(0)|0\rangle,
\end{eqnarray}
with light-cone directed $\zeta_{\mu}$ satisfying the nonsingular
condition $\eta^{\mu\nu}\zeta_{\mu}\zeta_{\nu}=0$
for a (super) current in a flat background,
we can choose the gauge parameters in (\ref{eq:puregauge_b})
and (\ref{eq:puregauge_s}) as
\begin{eqnarray}
&&\Lambda=U_1^{\dagger} U_1 i\zeta_{\mu}X^{\mu}(0)|0\rangle\,,
\label{eq:gauge_ex_b}\\
&&\Lambda_2=U_1^{\dagger} U_1\xi 
i\zeta_{\mu}X^{\mu}(0)|0\rangle\,,
~~~
\Lambda_1=U_1^{\dagger} U_1
c\xi\partial\xi e^{-2\phi}i\zeta_{\mu} X^{\mu}(0)|0\rangle\,,
\label{eq:gauge_ex_s}
\end{eqnarray}
{\it if we regard $X^{\mu}$ as a dimension $0$ primary field}.
In this case,
it is seen that the induced field redefinitions (\ref{eq:redef_b})
and (\ref{eq:redef_s}) involve the zero mode of $X^{\mu}$,
which implies the effect of the background Wilson line,
as investigated in Refs.~\citen{TT_mar, TT_tach} and \citen{KT_super}.

We should note that the string fields $\Lambda$ in (\ref{eq:gauge_ex_b})
and $\Lambda_1$ and $\Lambda_2$ in (\ref{eq:gauge_ex_s})
are not well-defined when some directions are compactified,
because they include $X^{\mu}$ instead of $\partial X^{\mu}$.
In this case, the gauge parameter string fields
$U^{(\alpha,\beta)}$ in (\ref{eq:pure_gauge_b}) and 
$U_{(i)}^{(\alpha,\beta)}$ and $V_{(i)}^{(\alpha,\beta)}$ in
(\ref{eq:pure_gauge_s}) are not well-defined. 
This implies that the solutions 
$\Psi^{(\alpha,\beta)}$ and  $\Phi^{(\alpha,\beta)}_{(i)}$,
which are independent of the zero mode of $X^{\mu}$,
are not globally pure gauge forms.

\section{Concluding remarks
\label{sec:discussion}}

In this paper, we have examined a class of
solutions to the equations of motion:
$\Psi^{(\alpha,\beta)}$ appearing in (\ref{eq:b_sol_gen})
in bosonic string field theory
and  $\Phi^{(\alpha,\beta)}_{(i)}$  appearing in
(\ref{eq:Phi_1})--(\ref{eq:Phi_4})
in superstring field theory.
The former is a variant of solutions given in Refs.~\citen{Schnabl_tach, 
Schnabl:2007az, Kiermaier:2007ba}, and the latter is a variant of 
solutions given in Refs.~\citen{Erler_super} and \citen{Okawa_super}.
Both solutions are generated from simpler ones, nemely, those in
(\ref{eq:psi_trivial}) and (\ref{eq:hatphi_super}), for example,
which are constructed from the identity state,
by using a commutative monoid $\{P_{\alpha}\}_{\alpha\ge 0}$ and
the associated $A^{(\gamma)}$ and $\hat{A}^{(\gamma)}$.
We have investigated gauge transformations among our solutions,
their (formal) pure gauge forms and induced string field redefinitions.
Using the family of wedge states as $\{P_{\alpha}\}_{\alpha\ge 0}$,
we have performed some explicit computations.

In \S\ref{sec:gauge}, we found finite gauge transformations 
that relate our solutions $\Psi^{(\alpha,\beta)}$ and
$\Phi^{(\alpha,\beta)}_{(i)}$ to the
simple solutions $\hat\psi$ and $\hat\phi$,
respectively, using path-ordered forms.
In this sense, our solutions constructed from  
nonsingular (super) currents, including those given in
Refs.~\citen{Schnabl:2007az} and \citen{Kiermaier:2007ba} for bosonic
strings and Refs.~\citen{Erler_super} and \citen{Okawa_super} 
for superstrings, are all {\it formally} gauge equivalent to simple
solutions based on the identity state.
\\

As mentioned in Footnote \ref{footnote:zeze},
$\Psi^{(\alpha,\beta)}$ appearing in (\ref{eq:b_sol_gen})
represents a solution if $\hat{\psi}$ is a solution to 
the equation of motion (\ref{eq:EOM_bos}) in bosonic string field
theory. This implies that $\hat{\psi}\mapsto \Psi^{(\alpha,\beta)}$ is 
a map from one solution to another. Let us denote this
transformation by $\Psi^{(\alpha,\beta)}(\hat{\psi})$.
We note that the composition of this transformation forms 
a commutative monoid, because the relations
$\Psi^{(\alpha,\beta)}(\Psi^{(\alpha',\beta')}(\hat{\psi}))
=\Psi^{(\alpha+\alpha',\beta+\beta')}(\hat{\psi})
=\Psi^{(\alpha',\beta')}(\Psi^{(\alpha,\beta)}(\hat{\psi}))$ and 
$\Psi^{(0,0)}(\hat{\psi})=\hat{\psi}$ hold
if we use the family of wedge states as $\{P_{\alpha}\}_{\alpha\ge 0}$
and the associated $A^{(\gamma)}$ given in (\ref{eq:Agamma_ex}).
As an application of this transformation $\Psi^{(\alpha,\beta)}(\cdot)$,
we can obtain new solutions from the Takahashi-Tanimoto marginal 
solution $\Psi^{\rm TT}_{\rm m}$ and scalar
solution $\Psi^{\rm TT}_{\rm s}$ derived in Ref.~\citen{TT_tach},
at least naively. The generated solutions are not based on
the identity state if we take $\alpha,\beta>0$, unlike the original
ones. Therefore, we conjecture that we can evaluate the action at
$\Psi^{(\alpha,\beta)}(\Psi^{\rm TT}_{\rm s})$ directly, 
in principle.
At least formally,
$\Psi^{(\alpha,\beta)}(\Psi^{\rm TT}_{\rm m})$ gives 
a solution for a general marginal deformation $\zeta_aJ^a$ with singular
OPE, namely, $\zeta_a\zeta_bg^{ab}\ne 0$ with (\ref{eq:JJ_OPE}),
because $\Psi^{\rm TT}_{\rm m}$ is constructed using
general currents \cite{TT_mar, TT_tach, KT_super}.
This might be an alternative approach to constructing the
solutions for marginal deformations with general currents
given in Refs.~\citen{Kiermaier:2007ba} and \citen{FKP}.

Similarly, comparing (\ref{eq:pure_gauge_b}) with (\ref{eq:Phi_2}),
we have found a map from one general solution to another
by modifying  $\Phi^{(\alpha,\beta)}_{(2)}$
appropriately in superstring field theory.
Specifically, if $\hat{\phi}$ satisfies the equation of motion
$\eta_0(e^{-\hat{\phi}}Q_{\rm B}e^{\hat{\phi}})=0$,
then $\Phi^{(\alpha,\beta)}$ given by
\begin{eqnarray}
\Phi^{(\alpha,\beta)}
&=&\log(1+P_{\alpha}*\hat f*P_{\beta})\,,
~~~~~\hat f
=(e^{\hat{\phi}}-I)*\frac{1}{1
-\eta_0\hat{A}^{(\alpha+\beta)}*e^{-\hat\phi}Q_{\rm B}e^{\hat{\phi}}}
\end{eqnarray}
also does: i.e.,
$\eta_0(e^{-\Phi^{(\alpha,\beta)}}
Q_{\rm B}e^{\Phi^{(\alpha,\beta)}})=0$.
Using this formula, we should be able to generate new solutions that
are not based on the identity state for $\alpha,\beta>0$,
using general supercurrents $\zeta_a{\bf J}^a(z,\theta)$ with 
$\zeta_a\zeta_b\Omega^{ab}\ne 0$
(\ref{eq:super_c_1})--(\ref{eq:super_c_3})
from identity-based solutions obtained in Ref.~\citen{KT_super}.
It is an interesting problem to closely examine such generated solutions
\cite{KMZ}. \\

It is an important problem to define the regularity 
of string fields more carefully and examine 
that of our generated solutions 
and the gauge parameters that relate them, because we have used some
formal properties of the star product and formally evaluated
 infinite summations.
In particular, such a problem seems to be very important to 
studying the cohomology around solutions. 
(See Refs.~\citen{Kishimoto:2002xi} and \citen{ES}, for example.)


\section*{Acknowledgements}

We are grateful to T.~Takahashi for his collaboration at various stages
of this work.
We would like to thank Y.~Okawa and S.~Zeze for valuable comments.
Discussions during the workshops ``Komaba 2007'' and 
``KEK Theory Workshop 2007'' were also useful.
I.~K. would like to thank the members of the
National Center for Theoretical Science
(North) in Taipei, where this work was completed,
for their hospitality.
I.~K. is supported in part by the Special Postdoctoral Researchers
Program at RIKEN and a Grant-in-Aid for Young
Scientists (\#19740155) from the Ministry of Education,
Culture, Sports, Science and Technology of Japan.

%

\end{document}